\documentclass[a4paper,aps,superscriptaddress,reprint,showpacs]{revtex4-1}
\usepackage{amsmath, amssymb}
\usepackage{bm, braket}
\usepackage{graphicx}
\usepackage[colorlinks=true,linkcolor=blue,citecolor=blue,urlcolor=blue]{hyperref}
\usepackage{xspace}
\usepackage{color}
\usepackage{braket}
\usepackage{bm}
\usepackage[all, warning]{onlyamsmath}
\usepackage{multirow}
\usepackage{here}
\usepackage{ulem}
\usepackage{xcolor}

\newcommand{\tb}[1]{#1}

\newcommand{\LCO}{La$_{2}$CuO$_{4}$\xspace}
\newcommand{\LSCO}{La$_{1.55}$Sr$_{0.45}$CuO$_{4}$\xspace}
\newcommand{\LSCOx}{La$_{2-x}$Sr$_{x}$CuO$_{4}$\xspace}
\newcommand{\bmR}{\bm{R}\xspace}
\newcommand{\bmRp}{\bm{R}^\prime\xspace}

\newcommand{\add}[1]{#1}

\newcommand\del{\bgroup\markoverwith{\textcolor{magenta}{\rule[0.5ex]{3pt}{1pt}}}\ULon}

\begin{document}
\title{\textit{Ab initio} derivation of effective Hamiltonian for La$_{2}$CuO$_{4}$/La$_{1.55}$Sr$_{0.45}$CuO$_{4}$ heterostructure}

\author{Terumasa Tadano}
\affiliation{International Center for Young Scientists (ICYS), National Institute for Materials Science, Tsukuba, Ibaraki 305-0047, Japan}
\affiliation{Research and Services Division of Materials Data and Integrated System (MaDIS), National Institute for Materials Science, Tsukuba, Ibaraki 305-0047, Japan}

\author{Yusuke Nomura}
\affiliation{Department of Applied Physics, The University of Tokyo, 7-3-1 Hongo, Bunkyo-ku, Tokyo 113-8656, Japan}

\author{Masatoshi Imada}
\affiliation{Department of Applied Physics, The University of Tokyo, 7-3-1 Hongo, Bunkyo-ku, Tokyo 113-8656, Japan}

\begin{abstract}
We formulate a method of deriving effective low-energy Hamiltonian for nonperiodic systems such as interfaces for strongly correlated electron systems by extending conventional multi-scale {\it ab initio} scheme for correlated electrons (MACE). We apply the formalism to copper-oxide high $T_{\rm c}$ superconductors in an example of the interface between overdoped \LSCOx and Mott insulating \LCO recently realized experimentally. 
We show that the parameters of the $E_g$ Hamiltonian derived for the \LCO/\LSCO superlattice differ considerably from those for the bulk \LCO, particularly significant in the partially-screened Coulomb parameters and the level offset between the $d_{x^{2}-y^{2}}$ and $d_{z^{2}}$ orbitals, $\Delta E$. In addition, we investigate the effect of the lattice relaxation on the $E_g$ Hamiltonian by carefully comparing the parameters derived before and after the structure optimization. We find that the CuO$_{6}$ octahedra distort after the relaxation as a consequence of the Madelung potential difference between the insulator and metal sides, by which the layer dependence of the hopping and Coulomb parameters becomes more gradual than the unrelaxed case. Furthermore, the structure relaxation dramatically changes the $\Delta E$ value and the occupation number at the interface. This study not only evidences the importance of the ionic relaxation around interfaces but also provides a set of layer-dependent parameters of the $E_g$ Hamiltonian, which is expected to provide further insight into the interfacial superconductivity when solved with low-energy solvers.
\end{abstract}
\maketitle


\section{Introduction}

Interface is at a frontier of condensed matter research and properties and functions not attainable in bulk crystals are the subjects of recent extensive studies. Among all, superconductivity is one of the hottest topics, where interface atomic layers often show properties superior to the bulk in terms of the critical temperature and its stability. Examples are interfaces of copper-oxide superconductors~\cite{Zhou:2010hg,Logvenov,Wu:2013cz,PhysRevB.89.184509} and the iron-based superconductors such as FeSe grown on the substrate such as SrTiO$_3$~\cite{FeSe,He:2013aa,0256-307X-29-3-037402}.  

Among all, recent experimental realization of the pinning of the critical temperature for the interface between overdoped \LSCOx and Mott insulating \LCO~\cite{Wu:2013cz} has inspired several theoretical studies~\cite{Misawa2016SciAdv}.  Experimentally, $T_{\rm c}$ is pinned at 40 K, which is the highest critical temperature of the bulk even when the doping concentration is varied in a wide range of $0.2<x<0.5$ in the overdoped side \LSCOx, indicating stable self-optimization of the superconductivity by the interface structure.

However, mechanisms of such fascinating phenomena at interfaces are difficult to identify experimentally in general, because interfaces give only tiny (negligible) contribution to thermodynamic quantities. In addition, the surface sensitive probes such as photoemission and scanning tunnel microscope spectroscopies are not suitable in contrast to surfaces. Therefore, even the lattice constant is hardly determined.  

Given this situation, the role of first principles studies, which can predict lattice parameters and atomic positions, 
becomes more important.
Furthermore, in the case of interfaces in strongly correlated electron systems, 
we need to take into account the effect of electron correlations properly. 
To study the correlation effect from first principles, 
derivation of low-energy effective Hamiltonian describing the degrees of freedom near the Fermi level 
is useful~\cite{ImadaMiyake:2010,Kotliar:RMP2006}. 
However, the lack of the periodicity makes calculations challenging and so far there exist 
only few applications to interface systems~\cite{Hirayama:JPSJ2012}.

To derive effective low-energy Hamiltonians for interfaces in strongly correlated electron systems from first prinpiples,
we need to extend the formalism developed for the bulk systems~\cite{ImadaMiyake:2010}. In the formalism for the bulk periodic systems, the low-energy effective Hamiltonians are
derived without any adjustable parameters based on the multi-scale {\it ab initio} scheme for correlated electrons (MACE) after taking the partial trace summation for the degrees of freedom out of the target low-energy space. The partial trace summation is taken by following either the constrained random phase approximation (cRPA) or the constrained GW (cGW) approximation~\cite{Hirayama:2013,Hirayama:2017}. 
All of these formalisms at the moment almost always use the experimental lattice structure and parameters of materials, although the lattice relaxation and optimization can be used without relying on the experimental values if the materials are not available and one wishes to design.
However, for the interface, even for the experimentally available systems, the lattice parameters are in many cases not available.  
Therefore, for the interface calculation, we need first to relax and optimize the lattice structure and predict the precise lattice parameters. 

In this paper, we propose a formalism by extending MACE to develop a scheme suitable for nonperiodic systems such as interfaces first by implementing the lattice relaxation near the interface.  This procedure is next combined with the conventional MACE treatment.    
To show its performance, we take an example of the interface between overdoped La$_{2-x}$Sr$_x$CuO$_4$ and Mott insulating La$_2$CuO$_4$.  
To gain insight beyond the previous work~\cite{Misawa2016SciAdv}, we examine the effect of lattice relaxation, which was not studied before.

We derive the two-band effective Hamiltonian consisting mainly of the antibonding band formed from Cu $3d_{x^2-y^2}$ and O $2p_{\sigma}$ orbitals and the Cu $3d_{3z^2-r^2}$ ($3d_{z^2}$) band for La compounds by extending the bulk studies~\cite{Hirayama:2018,Jang:2016aa}. The reason why we do not employ the one-band Hamiltonian is that the above two bands are severely hybridized in the case of La$_2$CuO$_4$~\cite{Matt:2018aa}.

The structure of the paper is the following: In Sec.~\ref{Sec:method}, we present the method. 
In Secs. \ref{Sec:result_bulk} and \ref{Sec:result_hetero}, we show the results of lattice relaxation and derived two-band Hamiltonian parameters for bulk and interface systems, respectively. 
Secs. \ref{Sec:discussion} and \ref{Sec:conclusion} are devoted to discussion and conclusion of the paper.

\section{Method}
\label{Sec:method}
We derive \textit{ab initio} low-energy effective Hamiltonians for \LCO and its heterostructures by employing the cRPA scheme~\cite{Aryasetiawan:2004kq} and the maximally localized Wannier function (MLWF) method~\cite{Marzari:2012eu} on top of density functional theory (DFT) calculations.

We first choose the target low-energy subspace ($d$ subspace) to construct the effective Hamiltonian for $d$-subspace electrons.
In the case of \LCO, we adopt the two-band Hamiltonian comprising the \add{Cu $3d_{x^{2}-y^{2}}$-like and $3d_{z^{2}}$-like orbitals near the Fermi level. Unlike the case of HgBa$_{2}$CuO$_{4}$ where the one-band Hamiltonian should be a good minimum model, the two $3d$ orbitals are strongly entangled in \LCO. Therefore, the minimum effective Hamiltonian of \LCO should include at least two orbitals. Note that, in what follows, the ``Cu $3d_{x^{2}-y^{2}}$-like orbitals" are actually the antibonding orbital of the strongly hybridizing copper $3d_{x^{2}-y^{2}}$ and oxygen $2p_{\sigma}$ orbital.
}

The form of the effective two-orbital Hamiltonian is
\begin{align}
\mathcal{H} 
& = \sum_{\sigma}\sum_{\bmR\bmRp}\sum_{mn}t_{m\bmR n\bmRp}a_{m\bmR}^{\sigma\dagger}a_{n\bmRp}^{\sigma} - \sum_{\sigma \bmR m}\mu_{m} n_{m\bmR}^{\sigma} \notag \\
& + \frac{1}{2}\sum_{\sigma\rho}\sum_{\bmR\bmRp}\sum_{mn} \big\{ U_{m\bmR n\bmRp}a_{m\bmR}^{\sigma\dagger}a_{n\bmRp}^{\rho\dagger}a_{n\bmRp}^{\rho}a_{m\bmR}^{\sigma} \notag \\
& + J_{m\bmR n\bmRp}( a_{m\bmR}^{\sigma\dagger} a_{n\bmRp}^{\rho\dagger} a_{m\bmR}^{\rho}a_{n\bmRp}^{\sigma} + a_{m\bmR}^{\sigma\dagger} a_{m\bmR}^{\rho\dagger} a_{n\bmRp}^{\rho} a_{n\bmRp}^{\sigma})\big\},
\end{align}
where $a_{n\bmR}^{\sigma\dagger}$ ($a_{n\bmR}^{\sigma}$) is a creation (annihilation) operator of an electron with spin $\sigma$ in the $n$th Wannier orbital located in the unit cell at position $\bm{R}$, $t_{m\bmR n\bmRp}$ is the hopping parameter, $\mu_{m}$ is the orbital-dependent on-site potential, $n_{m\bmR}^{\sigma}=a_{m\bmR}^{\sigma\dagger}a_{m\bmR}^{\sigma}$, and $U_{m\bmR n\bmRp}$ and $J_{m\bmR n\bmRp}$ are effective Coulomb and exchange interactions, respectively. 
Let $\mathcal{H}_{\mathrm{KS}}$ denote the Kohn-Sham (KS) Hamiltonian of DFT calculation and $W_{\rm eff}$ be the partially-screened Coulomb interaction, the hopping and Coulomb parameters are expressed as follows:
\begin{align}
&(1 - \delta_{mn}\delta_{\bmR\bmRp}) t_{m\bmR n\bmRp} - \delta_{mn}\delta_{\bmR\bmRp} \mu_{m} \notag \\ 
& \hspace{30mm} = \braket{\phi_{m\bmR}|\mathcal{H}_{\mathrm{KS}}|\phi_{n\bmRp}}, \\
&U_{m\bmR n\bmRp} = \braket{\phi_{m\bmR}\phi_{n\bmRp}| W_{\rm eff} |\phi_{m\bmR}\phi_{n\bmRp}}, 
\label{Eq:UcRPA} \\
&J_{m\bmR n\bmRp} = \braket{\phi_{m\bmR}\phi_{n\bmRp}| W_{\rm eff} |\phi_{n\bmRp}\phi_{m\bmR}},
\label{Eq:JcRPA}
\end{align}
with $\ket{\phi_{m\bmR}}= a_{m\bmR}^{\dagger}\ket{0}$ and $\delta_{ij}$ being the Kronecker delta.

In the calculation of the effective interaction parameters in Eqs.~(\ref{Eq:UcRPA}) and (\ref{Eq:JcRPA}), 
we exclude the screening contribution associated with the $d$-$d$ polarization processes within the $E_g$ manifold of the $3d$ electron subspace (namely, $d$ subspace) and use  
the partially screened interactions $W_{\rm eff}$: The $d$-subspace screening contribution is considered when we analyze the low-energy Hamiltonians, therefore, we need to exclude it in deriving $U_{m\bmR n\bmRp}$ and $J_{m\bmR n\bmRp}$ 
to avoid the double counting of the screening~\cite{Aryasetiawan:2004kq}.  
Then $W_{\rm eff}$ is given by 
\begin{align}
 W_{\rm eff} =   ( 1- v P_r  )^{-1} v  
\end{align}
where $P_r$ is given by $P_r = P - P_d$ with the full polarization $P$ and $d$-subspace polarization $P_{d}$, and 
$v$ is the bare Coulomb interaction. 

When the $d$ subspace is not isolated from the high-energy subspace ($r$ subspace), which is usually the case in cuprates, it is necessary to handle the entanglement to construct the $d$-subspace polarization $P_{d}$~\cite{Miyake:2009kx,Sasoglu:2011ch}. 
In this study, we employ the simple approach of Ref.~\onlinecite{Sasoglu:2011ch}, where the matrix element of $P_{d}$ in the plane-wave basis is given as 
\begin{align}
P^{d}_{\bm{G}\bm{G}'}(\bm{q},\omega) &= \sum_{\bm{k},\alpha,\beta}\rho_{\alpha\beta\bm{k}\bm{q}}^{*}(\bm{G})\rho_{\alpha\beta\bm{k}\bm{q}}(\bm{G}')  F^{d}_{\alpha\beta}(\bm{k},\bm{q})  \label{eq:Pd} \notag \\
& \hspace{3mm} \times \left[  \frac{1}{\omega + G_{\alpha\beta} (\bm{k},\bm{q})} -  \frac{1}{\omega - G_{\alpha\beta} (\bm{k},\bm{q})}  \right], \\
\rho_{\alpha\beta\bm{k}\bm{q}}(\bm{G}) &= \braket{\psi_{\alpha\bm{k}+\bm{q}} | e^{\mathrm{i}(\bm{q}+\bm{G})\cdot\bm{r}}|\psi_{\beta\bm{k}}}, \\
F^{d}_{\alpha\beta}(\bm{k},\bm{q}) &= \theta (\epsilon_{\alpha,\bm{k}+\bm{q}} - \epsilon_{\mathrm{F}})\theta(\epsilon_{\mathrm{F}}-\epsilon_{\beta,\bm{k}}) \notag \\ 
&\hspace{35mm} \times w_{\alpha\bm{k}+\bm{q}}w_{\beta\bm{k}}, \label{eq:F_pol}\\
G_{\alpha\beta}(\bm{k},\bm{q}) &= \epsilon_{\alpha,\bm{k}+\bm{q}} - \epsilon_{\beta, \bm{k}} + \mathrm{i}\delta.
\end{align}
Here, $\epsilon_{\alpha\bm{k}}$ and $\ket{\psi_{\alpha\bm{k}}}$ are the KS eigenvalue and wavefunction of the $\alpha$th band at the momentum $\bm{k}$, respectively, $\bm{G}$ is the reciprocal lattice vector, $\epsilon_{\mathrm{F}}$ is the Fermi level, $\theta(x)$ is the Heaviside step function, and $\delta$ is a small negative value.
The term $w_{\alpha\bm{k}}$ in Eq.~(\ref{eq:F_pol}) is the weight of the target Wannier orbital defined as
\begin{equation}
w_{\alpha\bm{k}} = \sum_{m} |U_{\alpha m}^{(\bm{k})}|^{2},
\end{equation}
with $U_{\alpha m}^{(\bm{k})}$ being a unitary matrix that transforms $\psi_{\alpha\bm{k}}(\bm{r})$ into $\phi_{m\bm{R}}(\bm{r})$ as 
\begin{align}
  \phi_{m\bm{R}}(\bm{r}) &= \frac{1}{\sqrt{N}}\sum_{\bm{k}} e^{-\mathrm{i}\bm{k}\cdot\bm{R}} \psi_{m\bm{k}}^{(w)} \notag \\
  & = \frac{1}{\sqrt{N}}\sum_{\bm{k}} e^{-\mathrm{i}\bm{k}\cdot\bm{R}}\sum_{\alpha} U_{\alpha m}^{(\bm{k})}\psi_{\alpha\bm{k}}(\bm{r}),
\end{align}
where $\psi_{m\bm{k}}^{(w)}$ is the Wannier-gauge Bloch wavefunction.
Hence, $w_{\alpha\bm{k}}$ can be obtained straightforwardly from the results of the MLWF calculation. 
If the KS wavefunction $\ket{\psi_{\alpha\bm{k}}}$ is strictly represented only within the subspace of the $d$ electrons, we obtain $w_{\alpha\bm{k}}=1$. Therefore, the term $w_{\alpha\bm{k}+\bm{q}}w_{\beta\bm{k}}$ in $F^{d}_{\alpha\beta}(\bm{k},\bm{q})$ becomes exactly one when both the virtual state $\ket{\psi_{\alpha\bm{k}+\bm{q}}}$ and the occupied state $\ket{\psi_{\beta\bm{k}}}$ belong to the $d$ subspace, and the $d$-$d$ transition in $P^{r}_{\bm{G}\bm{G}'}(\bm{q},\omega)= P_{\bm{G}\bm{G}'}(\bm{q},\omega)- P^{d}_{\bm{G}\bm{G}'}(\bm{q},\omega)$ is duly excluded.

Once we obtain $P_{d}$ by Eq.~(\ref{eq:Pd}), we can calculate the partially screened Coulomb parameters in Eqs. (\ref{Eq:UcRPA}) and (\ref{Eq:JcRPA}) as
\begin{align}
&U_{m\bm{0}n\bm{R}} = \frac{4\pi}{\Omega}\sum_{\bm{q}\bm{G}\bm{G}'}e^{-\mathrm{i}\bm{q}\cdot\bm{R}}\rho_{m\bm{q}}(\bm{G})\epsilon^{-1}_{\bm{G}\bm{G}'}(\bm{q},\omega=0)\rho_{n\bm{q}}^{*}(\bm{G}'), \label{eq:U_orig} \\
&J_{m\bm{0}n\bm{R}} = \frac{4\pi}{\Omega}\sum_{\bm{q}\bm{G}\bm{G}'}\rho_{mn\bm{R}\bm{q}}(\bm{G})\epsilon^{-1}_{\bm{G}\bm{G}'}(\bm{q},\omega=0)\rho_{mn\bm{R}\bm{q}}^{*}(\bm{G}'), 
\end{align}
where \add{$\Omega$ is the crystal cell volume,} $\epsilon^{-1}(\bm{q},\omega)$ is 
the inverse of the symmetric dielectric matrix, and $\rho_{m\bm{q}}(\bm{G})=\rho_{mm\bm{0}\bm{q}}(\bm{G})$ with $\rho_{mn\bm{R}\bm{q}}(\bm{G})$ being defined as follows:
\begin{equation}
\rho_{mn\bm{R}\bm{q}}(\bm{G}) = \frac{1}{N|\bm{q}+\bm{G}|}\sum_{\bm{k}}e^{-\mathrm{i}\bm{k}\cdot\bm{R}} \braket{\psi_{m\bm{k}+\bm{q}}^{(w)}|e^{\mathrm{i}(\bm{q}+\bm{G})\cdot\bm{r}}|\psi_{n\bm{k}}^{(w)}}. \label{eq:rho}
\end{equation}
The \add{symmetric} dielectric matrix is defined as 
\begin{equation}
\epsilon_{\bm{G}\bm{G}'}(\bm{q},\omega) = \delta_{\bm{G}\bm{G}'} - [v(\bm{q}+\bm{G})]^{\frac{1}{2}}P^{r}_{\bm{G}\bm{G}'}(\bm{q},\omega)[v(\bm{q}+\bm{G}')]^{\frac{1}{2}},
\end{equation}
where the bare Coulomb interaction $v(\bm{q})$ in reciprocal space is given by $v(\bm{q})=4\pi/\Omega|\bm{q}|^{2}$.

\section{ $E_{g}$ Hamiltonian of bulk systems }
\label{Sec:result_bulk}

To clarify specific properties at the LCO/LSCO interface, it is essential to first understand properties of the bulk systems including the doping-level dependence of the low-energy Hamiltonians. 
In this section, we carefully compare the $E_{g}$ Hamiltonians of the non-doped \LCO and the overdoped La$_{1.55}$Sr$_{0.45}$CuO$_{4}$.

All of the DFT calculations in this work were performed by using \textsc{Quantum ESPRESSO}~\cite{QE-2009}, which implements the plane-wave pseudopotential method. 
We employed the Pewdew--Burke--Ernzerhof (PBE) exchange-correlation potential~\cite{PhysRevLett.77.3865} and the optimized norm-conserving Vanderbilt (ONCV) pseudopotentials~\cite{Hamann:2013bq} from the SG15 table~\cite{Schlipf:2015fn}.
The kinetic-energy cutoff was set to 100 Ry.
We employed the tetragonal structure of \LSCOx (space group: $I4/mmm$), where the Sr doping was modeled by the virtual crystal approximation (VCA).

\begin{figure}[tb]
  \centering
  \includegraphics[width=5.5cm, clip]{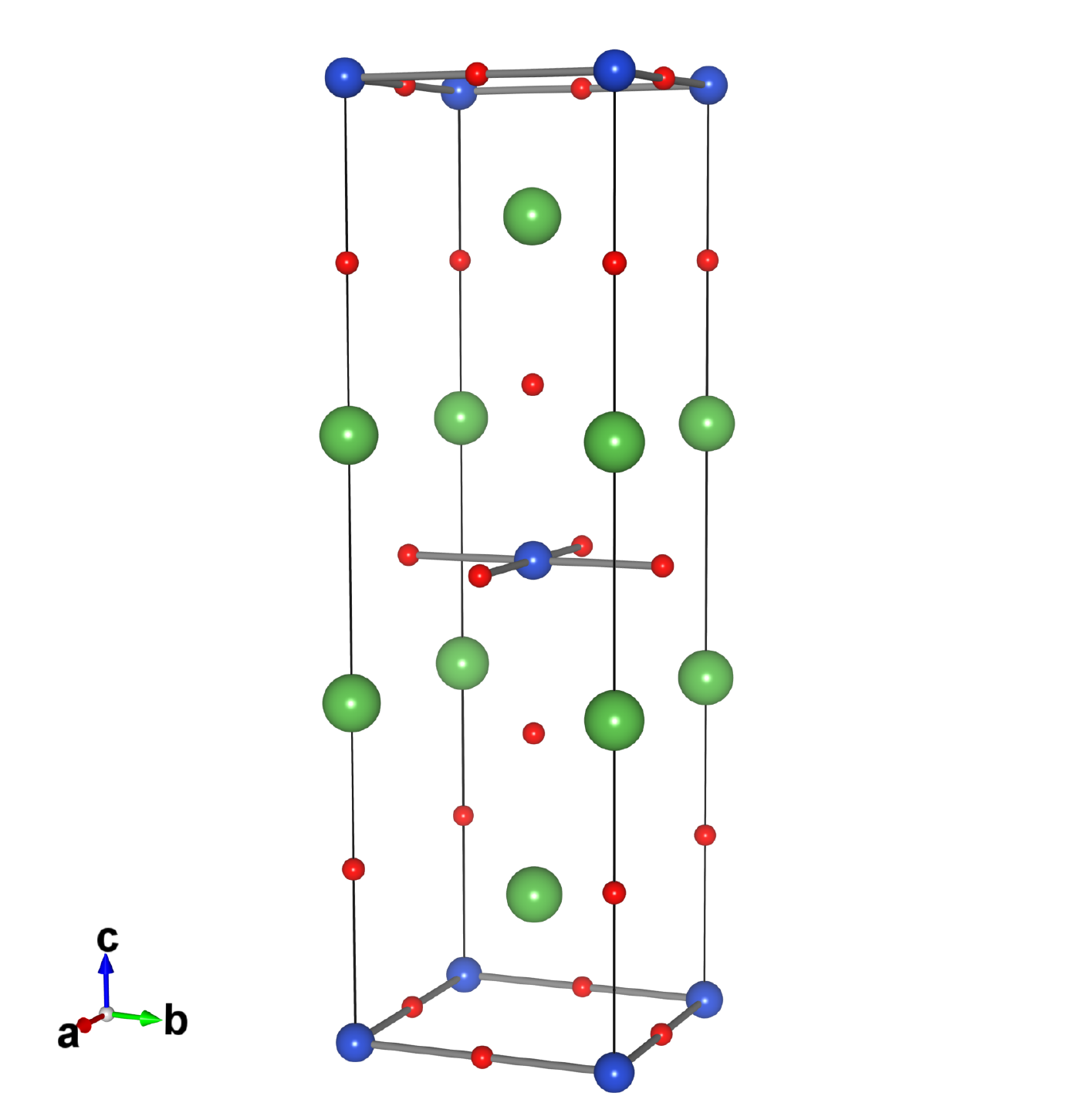}
  \caption{(color online) Crystal structure of the conventional unit cell of the tetragonal \LCO. The unit cell contains 2 CuO$_2$ layers.  Green, blue, and red spheres represent La, Cu, and O atoms, respectively (created with \textsc{vesta}~\cite{Momma:db5098}).}
  \label{fig:bulk_structure}
\end{figure}

When constructing the low-energy Hamiltonians of the bulk systems, we employed the conventional unit cell (see Fig.~\ref{fig:bulk_structure}) because it is more convenient than the primitive cell to apply consistent Wannierization parameters, particularly the frozen window, between the bulk and heterostructure systems. In the conventional unit cell calculations, we employed the 8$\times$8$\times$4 $\bm{k}$ points for the Brillouin zone (BZ) integration with the smearing width of 0.02 Ry. 

\subsection{Structural properties}

First, we fully relaxed the tetragonal structure of \LSCOx at different doping levels $x$ \add{until both the force convergence criteria $|F| < 10^{-6}$ Ry$\cdot$Bohr$^{-1}$ and the stress convergence criteria $|\sigma| < 0.5$ kbar are satisfied}. The results are compared with the available experimental data in Table~\ref{table:bulk_structure}. According to the paper of Radaelli \textit{et al.}~\cite{1994PhRvB..49.4163R}, the tetragonal phase of \LSCOx is stable at 0 K only when $x > 0.21$, below which the orthorhombic phase becomes the most stable at low temperatures. Nonetheless, it should be reasonable to use the tetragonal structure for the derivation of the effective Hamiltonian because no discontinuity in the $T_{\mathrm{c}}$ has been observed at the tetragonal-to-orthorhombic phase transition~\cite{1994PhRvB..49.4163R}. The in-plane lattice constants of the orthorhombic phase are approximately equal to  $\sqrt{2} a_{\mathrm{tetra}}$. Therefore, we show $\sqrt{2} a_{\mathrm{tetra}}$ instead of $a_{\mathrm{tetra}}$ in Table~\ref{table:bulk_structure} for the purpose of comparison.

\begin{table}[bt]
\caption{Comparison of calculated (calc.) and experimental (expt.) structural parameters of La$_{2-x}$Sr$_{x}$CuO$_{4}$. $h_{\mathrm{O}}$ is the distance between a Cu atom and its apical oxygen site. 
The calculations are performed for four different structures. 
We assign the labels for the undoped cases ($x=0$) as N1 and N2 and heavily doped cases ($x=0.45$) as D1 and D2.
In N1 and D1, the lattice constants and internal atomic coordinates are fully relaxed, whereas in N2 and D2 the planer lattice constant is fixed to match that of the substrate 
LaSrAlO$_{4}$ (see the main text for detail).
For the tetragonal phase, $a$ value in the table are $\sqrt{2} a_{\mathrm{tetra}}$ where $a_{\mathrm{tetra}}$ is the lattice constant of the tetragonal unit cell, and $b$ value is left blank.}
\label{table:bulk_structure}
\begin{ruledtabular}
\begin{tabular}{cccccc}
 Label & $x$ & $a$ (\AA) & $b$ (\AA) & $c$ (\AA) & $h_{\mathrm{O}}$ (\AA) \\
 \hline
Calc. (fully relaxed) \\
 N1  & 0.00 & 5.398 &  & 13.176 & 2.450  \\
  & 0.15 & 5.383 &  & 13.247 & 2.420 \\
  & 0.30 & 5.381 &  & 13.237 & 2.390 \\
 D1 & 0.45 & 5.388 &  & 13.186 & 2.346 \\ \\
 \multicolumn{3}{l}{ Calc. (relax $c$ with $a_{\mathrm{tetra}} = a_{\mathrm{LSAO}}$)} \\ 
 N2 & 0.00 & 5.310 &  & 13.354 & 2.481 \\ 
 D2 & 0.45 & 5.310 &  & 13.377 & 2.393 \\ \hline
  \multicolumn{2}{l}{Expt. (Ref.~\onlinecite{1994PhRvB..49.4163R}, 10 K)} \\
  & 0.00 & 5.335 & 5.415 & 13.117 &  2.420 \\
  & 0.15 & 5.325 & 5.349 & 13.197 & 2.414 \\
  & 0.30 & 5.312 &       & 13.228 & 2.390 \\
\end{tabular}
\end{ruledtabular}
\end{table}

It is observed in the table that our DFT results based on the VCA reasonably well reproduce the experimental cell parameters within 1\% error as well as the trend of the doping-level dependence, thus validating the reliability of the VCA. 
In Table~\ref{table:bulk_structure}, we also show the optimized values of the out-of-plane lattice constant $c$ with $a$ and $b$ being fixed to those of LaSrAlO$_{4}$ (LSAO), which was used as the substrate for growing the LCO/LSCO bilayer thin film~\cite{Zhou:2010hg}. Since the in-plane lattice constants of LSAO is slightly smaller than those of LCO and LSCO, the LSAO substrate induces the compressive strain along the $a$ and $b$ axes, leading to a slightly larger $c$ value due to positive Poisson ratio.

\subsection{Construction of MLWFs} 

Second, we construct the MLWFs of \LCO and \LSCO to see the doping-level and strain dependence of the hopping parameters and the Coulomb interaction. Since the target $E_{g}$ orbitals are not isolated, the resulting MLWFs are rather sensitive to the chosen range of the energy window. 
In this study, we set the outer window by band index. Here, the outer window specifies the Hilbert subspace, within which the MLWFs are constructed. We included 8 valence bands per CuO$_{2}$ layer in the outer window (see Fig.~\ref{fig:bulk_band}), which was the narrowest window setting to match the MLWF band structures with the KS ones around the Fermi level both for the non-doped and doped systems. Also, the frozen window~\cite{Marzari:2012eu} was used to perfectly reproduce the original KS band of the $d_{x^{2}-y^{2}}$-like orbitals at the Fermi energy. The resulting two MLWFs are rather extended as shown in Fig.~\ref{fig:bulk_isosurface}. 

It is possible to derive more localized Wannier orbitals by including more valence bands in the outer window. To see the influence of the outer window range on the effective Hamiltonian, we also show the calculated parameters when we included 14 valence bands per CuO$_{2}$ layer as in Ref.~\cite{Hirayama:2018} in the supplemental material (SM). 
This condition still excludes the bonding and non-bonding states formed by the Cu $d_{x^{2}-y^{2}}$ and in-plane O $p_{\sigma}$ orbitals from the energy window
but newly includes the bonding state formed by the Cu $d_{z^{2}}$ and apical-oxygen $p_{z}$ orbitals. Therefore, the resulting $d_{z^{2}}$-like MLWF becomes more localized than that of Fig.~\ref{fig:bulk_isosurface}(a), while the changes of the $d_{x^{2}-y^{2}}$ orbital shape and parameters are small (see SM).
As the two-band ($E_g$) description of the effective Hamiltonian, we believe that the choice of 8-band outer window is more appropriate, because the $d_{z^2}$ orbital is strongly hybridized with apical oxygen $p_z$ orbital. This strong hybridization is ignored if we employ the 14-band outer window.

\subsection{cRPA calculation} 

In the present cRPA calculations for bulk, 
we considered the particle-hole excitations within 150 bands (75 bands/f.u.) in calculating the polarization,
which corresponds to include 84 unoccupied bands (42 bands/f.u.) up to $\sim$ 21 eV above the Fermi level. The kinetic-energy cutoff for the polarization was set to 20 Ry, which was sufficiently large for the symmetric dielectric matrix to reach the large $|\bm{G}|$ limit of $\epsilon_{\bm{G}\bm{G}'}\approx \delta_{\bm{G}\bm{G}'}$.
The same computational conditions were employed for the cRPA calculation of interfaces.
The number of unoccupied bands was selected to make the cRPA calculation of the complex LCO/LSCO heterostructure feasible, but the screened Coulomb parameters, particularly the on-site Coulomb interaction, will be reduced further with increasing the number of bands. Fortunately, our setting still gives reasonably converged values of $U$. For example, in the N1 case, we obtained 3.92 eV for the on-site Coulomb interaction of the $d_{x^{2}-y^{2}}$ orbital. This value reduced to the almost converged result of 3.67 eV which were obtained with 500 bands. Therefore, the presented results for $U$ are within $\sim$ 7\% error (overestimate) from the converged values. 
Our $U$ value is in reasonable agreement with those of the previous cRPA~\cite{PhysRevB.91.125142,Jang:2016aa} and cGW~\cite{Hirayama:2018} studies as described in Appendix~\ref{Sec:cRPA_comparison}.

The values of effective Coulomb interactions computed by cRPA 
are determined by the shape (spatial spread) of MLWFs and the strength of screening. 
The doping and strain affect both of them. 
To see only the effect of the change in the shape of WLWFs, 
we also compute the Wannier matrix elements of the bare Coulomb interaction by replacing $W_{\rm eff}$ with $v$
in Eqs.~(\ref{Eq:UcRPA}) and (\ref{Eq:JcRPA}).

\subsection{Doping and strain dependence}
\label{Sec:bulk_doping_strain_dep}

\begin{figure}[tb]
  \centering
  \includegraphics[width=8.5cm, clip]{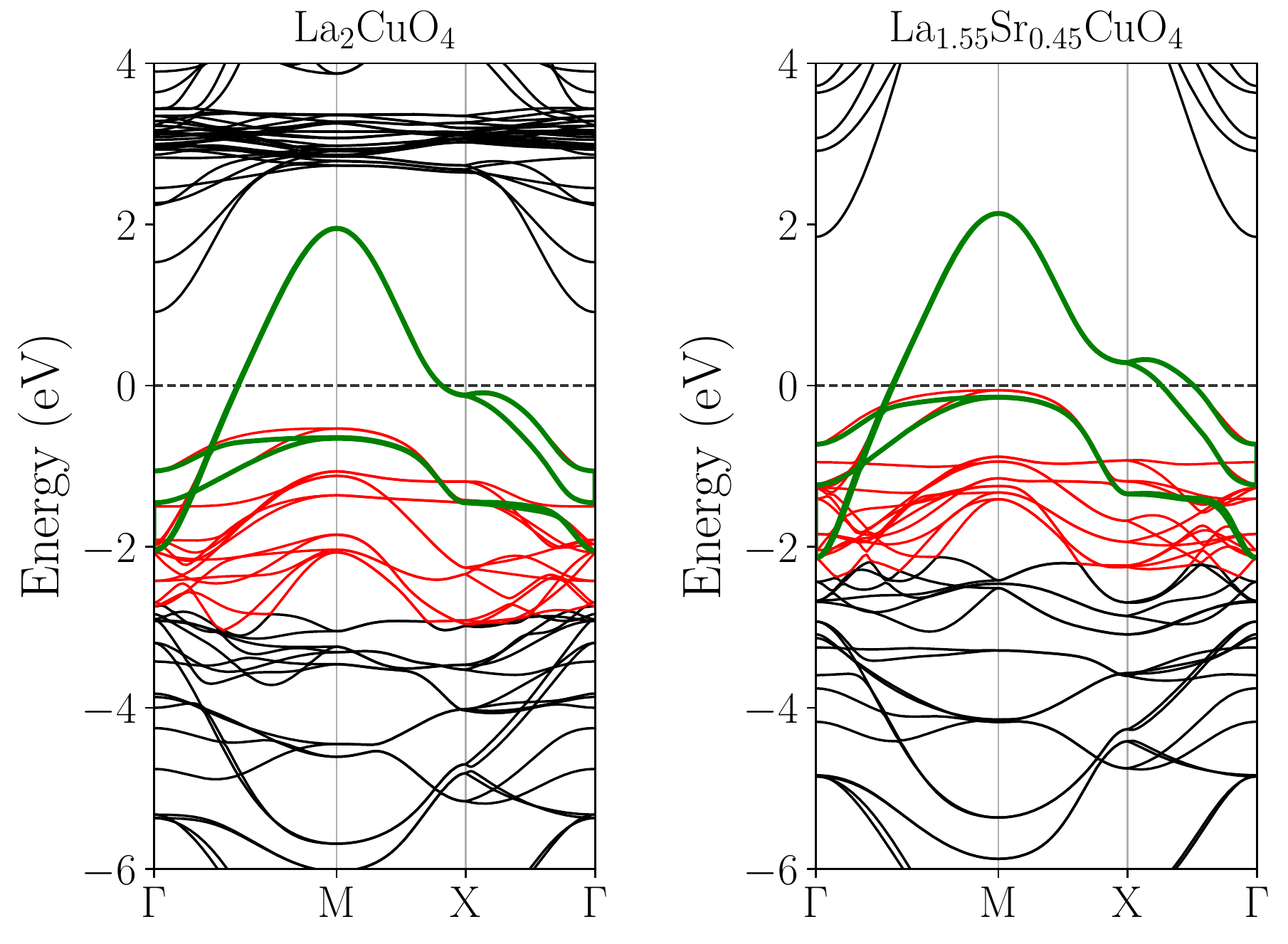}
  \caption{(color online) Band structure of \LSCOx ($x=0,0.45$) calculated with the conventional unit cell along the high-symmetry lines of the Brillouin zone (thin lines). 
  The energy is shown relative to the Fermi energy. The MLWF band structures of the $E_{g}$ Hamiltonian are also shown by green thick lines, which were constructed from the 16 KS bands (8 bands per CuO$_{2}$ layer) shown in red color.}
  \label{fig:bulk_band}
\end{figure}

\begin{figure}[tb]
  \centering
  \includegraphics[width=8.5cm, clip]{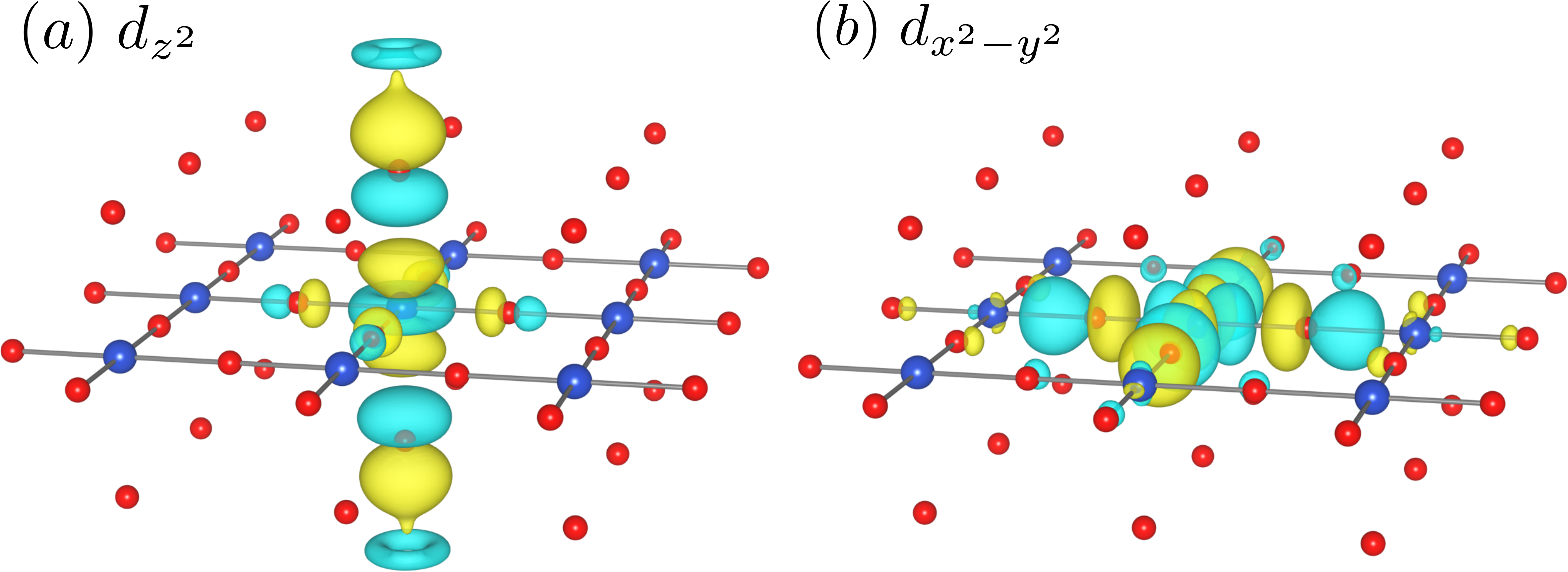}
  \caption{(color online) Isosurface of the constructed MLWFs of \LCO (isovalue = 0.03 au).
  The La atoms are not shown for a visualization purpose.}
  \label{fig:bulk_isosurface}
\end{figure}

Figure \ref{fig:bulk_band} compares the calculated KS and MLWF band structures of LCO and LSCO along the high-symmetry lines of the BZ. Here, the fully optimized structures (N1 and D1 in Table~\ref{table:bulk_structure}) are used. While the valence-band structures of the two systems are similar, a simple rigid-band picture seems insufficient for explaining the difference. To see the difference more clearly, we constructed atomic-like 17 MLWFs of LCO and LSCO from the isolated 17 valence bands consisting of the whole Cu $3d$ and O $2p$ manifold, 
and compared their onsite energy levels relative to the Fermi energy. 
We then observed that upon hole doping the energy levels of O $2p$ orbitals increased significantly, particularly for the apical oxygen, whereas those of Cu $3d$ orbitals changed only slightly. Hence, the hole doping by the chemical substitution reduces the energy-level difference of O $2p$ and Cu $3d$ orbitals. The observed shift of the O $2p$ energy levels can be attributed to the change of the electrostatic potential induced by the substitution, which mainly affects the O $2p$ orbitals of oxygen located near the La sites (see Fig.~\ref{fig:bulk_structure}).


The hole doping also affects the hopping and Coulomb parameters as can be seen in Table~\ref{table:bulk_parameter}.  
The significant change occurs for the partially screened on-site Coulomb parameters, whose reduction amounts to $\sim$ \add{21--31\%}. This reduction can be attributed to two different factors. One is the change of the spread of the MLWFs. Since the hole doping reduces the energy-level difference between the O $2p$ and Cu $3d$ orbitals as mentioned above, the $2p$-$3d$ hybridization becomes stronger, 
leading to more extended $E_{g}$ Wannier functions. 
The change of the MLWFs resulted in the 10--16\% reduction of the bare on-site Coulomb interactions. The other factor is the change of the screening strength. Upon hole doping, the energy levels of the O $2p$ orbitals become closer to the Fermi energy. This enhances the screening channel in $P_{r}$ and explains further reduction in $U$. 
As for the intraorbital hopping parameters, we see a slight increase upon doping.  
As a result of the decrease of the Coulomb interaction and the slight increase of the hopping parameters, the strength of correlation, which can be measured by the ratio $| U/t |$ for  $d_{x^{2}-y^{2}}$ orbital, changes significantly. 
$| U/t |$ changes from $\sim$ 8.5 (7.9) to $\sim$ 6.4 (5.8) for the fully-relaxed structure (relaxed structure with the constraint of $a = a_{\mathrm{LSAO}}$).

The level offset between the $d_{x^{2}-y^{2}}$ and $d_{z^{2}}$ orbitals $\Delta E = \mu_{x^{2}-y^{2}}-\mu_{z^{2}}$, which has been pointed out to be a key parameter to explain the material dependence of $T_{c}$~\cite{Sakakibara:2010jh,Sakakibara:2012br,Sakakibara:2012kr}, is also affected by hole doping. 
It decreases by 0.279 eV and 0.344 eV for the fully-relaxed structure and the relaxed structure with the constraint of $a = a_{\mathrm{LSAO}}$, respectively.

\begin{table*}[!h]
  \caption{Hopping and Coulomb parameters of the bulk $E_{g}$ Hamiltonians calculated with different doping level $x$ and structural parameters. The indices $m, n$ label the MLWFs; 0 and 1 correspond to  $d_{z^{2}}$ and $d_{x^{2}-y^{2}}$, respectively. The meanings of other parameters are the followings. $\Delta E$ : The on-site potential difference between the $d_{x^{2}-y^{2}}$ and $d_{z^{2}}$ orbitals; $t, t', t''$ : nearest, next-nearest, and second next-nearest hopping parameters
  on CuO$_2$ plane, respectively; $v, U$: on-site Coulomb interactions; $v_{n}, V_{n}$: nearest-neighbor Coulomb interactions; $J_{v}, J$ : exchange interactions; $|U/t|$ : scaled correlation strength for the $d_{x^{2}-y^{2}}$ orbital.}
  \label{table:bulk_parameter}
  \begin{ruledtabular}
    \begin{tabular}{ccrrrr}
                   &         & N1 ($x=0$) & N2 ($x=0$) & D1 ($x=0.45$) & D2 ($x=0.45$) \\ 
                   & $(m,n)$ &                   &                   &                      &                      \\ \hline
       $\Delta E$ 
                   &         &  $0.778$          & $ 1.034$          & $ 0.499$             & $ 0.690$  \\
               $t$ & $(0,0)$ & $-0.063$          & $-0.059$          & $-0.089$             & $-0.083$ \\
                   & $(0,1)$ & $ 0.170$          & $ 0.169$          & $ 0.204$             & $ 0.200$ \\
                   & $(1,1)$ & $-0.463$          & $-0.490$          & $-0.478$             & $-0.500$ \\
              $t'$ & $(0,0)$ & $-0.010$          & $-0.009$          & $-0.015$             & $-0.014$ \\
                   & $(0,1)$ & $ 0.000$          & $ 0.000$          & $ 0.000$             & $ 0.000$ \\
                   & $(1,1)$ & $ 0.092$          & $ 0.095$          & $ 0.096$             & $ 0.101$ \\
             $t''$ & $(0,0)$ & $-0.008$          & $-0.007$          & $-0.013$             & $-0.011$ \\
                   & $(0,1)$ & $ 0.027$          & $ 0.027$          & $ 0.033$             & $ 0.033$ \\
                   & $(1,1)$ & $-0.074$          & $-0.078$          & $-0.087$             & $-0.091$ \\
      Spread of the MLWF (\AA$^{2}$)
                   & $(0,0)$ & 4.00  &  3.99 &  4.72 &  4.85\\
                   & $(1,1)$ & 3.26  &  3.19 &  4.03 &  4.05 \\ \hline
      Bare Coulomb parameters \\
             $v$   & $(0,0)$ & 13.85 & 13.83 & 11.84 & 11.66\\ 
                   & $(0,1)$ & 12.55 & 12.53 & 10.74 & 10.53\\
                   & $(1,1)$ & 14.89 & 14.90 & 13.42 & 13.30\\
       $v_{\mathrm{n}}$
                   & $(0,0)$ & 3.33 & 3.37 & 3.27 & 3.28 \\
                   & $(0,1)$ & 3.63 & 3.68 & 3.56 & 3.59 \\
                   & $(1,1)$ & 4.10 & 4.17 & 4.10 & 4.16 \\
          $J_{v}$  & $(0,1)$ & 0.60 & 0.59 & 0.51 & 0.48 \\ \hline
      Partially screened Coulomb parameters \\    
           $U$  & $(0,0)$ & 3.94 & 3.91 & 3.02 & 2.89 \\
                   & $(0,1)$ & 2.63 & 2.60 & 1.93 & 1.80 \\
                   & $(1,1)$ & 3.92 & 3.89 & 3.07 & 2.92 \\
       $V_{\mathrm{n}}$ 
                   & $(0,0)$ & 0.63 & 0.64 & 0.61 & 0.59 \\
                   & $(0,1)$ & 0.72 & 0.72 & 0.69 & 0.66 \\
                   & $(1,1)$ & 0.85 & 0.86 & 0.83 & 0.81 \\
           $J$  & $(0,1)$ & 0.50 & 0.49 & 0.40 & 0.38     \\
           $|U/t|$ & $(1,1)$ & 8.47 & 7.94 &  6.42 & 5.84      
    \end{tabular}    
  \end{ruledtabular}  
\end{table*}

Next, we discuss the strain dependence. 
To see the effect of the compressive stress along the $ab$-plane induced by the LSAO substrate, we also calculated the hopping and Coulomb parameters with the N2 and D2 structures of Table~\ref{table:bulk_structure}.
As shown in Table~\ref{table:bulk_parameter}, the compressive stress changes the Coulomb parameters only slightly but significantly increases the level offset $\Delta E$ by $\sim$ 0.19--0.26 eV. This tendency agrees with the previous theoretical result~\cite{Sakakibara:2012kr}. 

We see that the doping and strain affect the parameters in the effective Hamiltonians. 
However, the changes of parameters upon doping and/or strain are usually neglected in previous studies. 
It is intriguing to study the superconducting amplitude and its competition with the charge inhomogeneity such as stripes by solving the present Hamiltonian using highly accurate low-energy solvers.

\section{E$_{g}$ Hamiltonian of heterostructures}
\label{Sec:result_hetero}

\begin{figure}
  \centering
  \includegraphics[width=7cm]{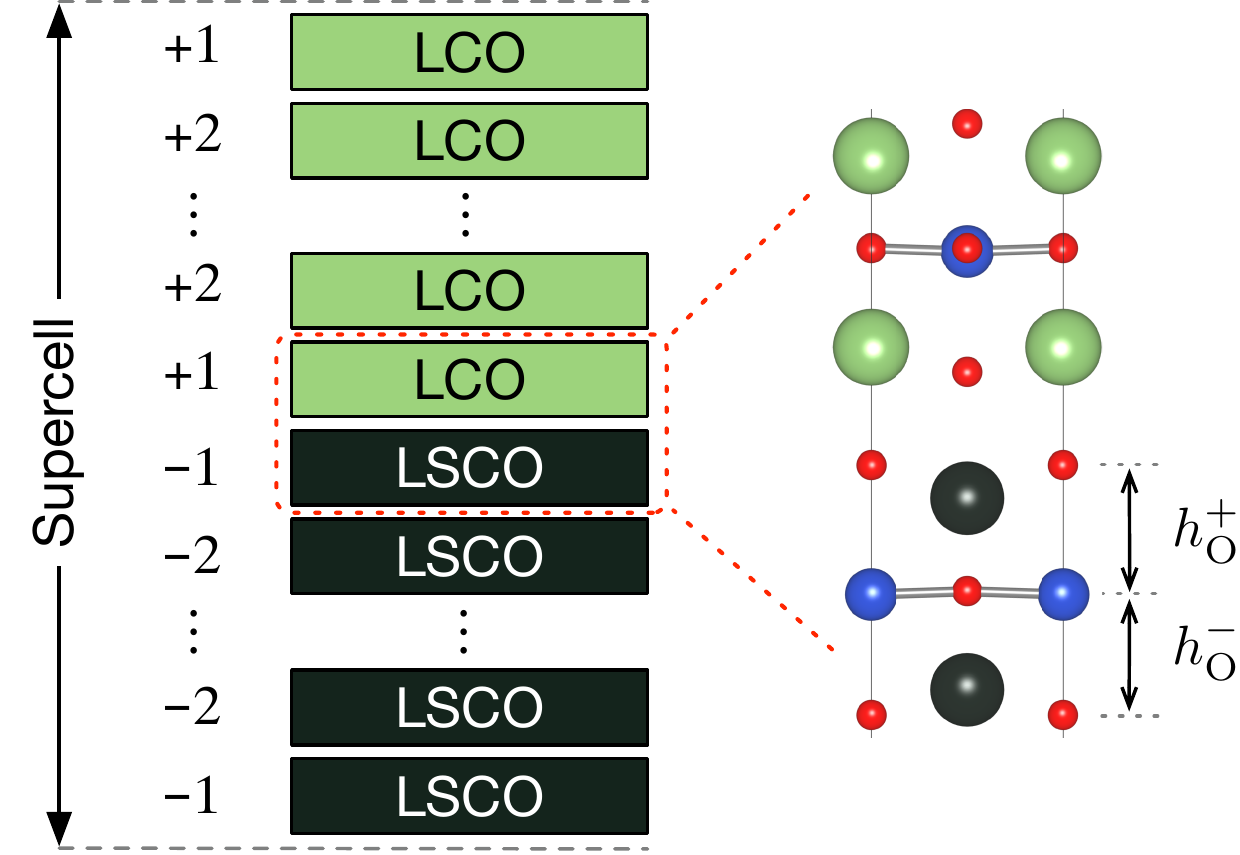}
  \caption{(color online) Schematic figure of the heterostructure of \LCO (LCO) and \LSCO (LSCO) employed in this study. The layer index ($\pm 1$, $\pm 2$, $\dots$) is assigned to inequivalent unit cells (CuO$_{2}$ layers). The atomic structure at the interface is also shown on the right hand side.}
  \label{fig:superlattice}
\end{figure}

To derive the $E_{g}$ Hamiltonian of the LCO/LSCO interface, we employ the superlattice (SL) structures schematically shown in Fig.~\ref{fig:superlattice}. The heterostructure was constructed by stacking tetragonal unit cells of LCO and LSCO along $c$ axis. For the LCO and LSCO structure units, we employed the N2 and D2 conventional cells in Table~\ref{table:bulk_parameter}, respectively.
Since each unit cell of LCO contains 7 atoms, the structural model of a $(M,N)$ superlattice contains 7$\times (M+N)$ atoms, where $M$ and $N$ are the numbers of unit cells in the LCO and LSCO regions, respectively.
Second, we performed DFT calculations of the $(M,N)$ superlattice and optimized the lattice constant along the $c$ axis as well as the internal coordinates, while the $a$ value was fixed to $a_{\mathrm{LSAO}}$.
Third, we constructed MLWFs of all Cu $E_{g}$ orbitals, corresponding to $2\times (M+N)$ total orbitals, and calculated the hopping, bare Coulomb, and partially-screened Coulomb parameters.
To see the convergence of the parameters with respect to the number of layers, we calculated hopping parameters for $(M,N) = $ (4,4), (6,6), and (8,8). The Coulomb parameters were calculated only for $(M,N) = $ (4,4) and (6,6) due to the computational limitations. In the structural optimization, we employed the 8$\times$8$\times$1 $\bm{k}$ points. In the subsequent cRPA calculations, the 8$\times$8$\times$2 $\bm{k}$-point mesh was employed in order to use the tetrahedron method for an accurate treatment of the summation over $\bm{k}$ in Eq.~(\ref{eq:Pd}). 
In the calculations of SLs, the Brillouin zone becomes highly anisotropic. In this case, we found that the original definition of Eq.~(\ref{eq:rho}) needs to be modified to perform the stable cRPA calculation as is discussed in detail in Appendix~\ref{Sec:Averaging}.  
Our modified $\tilde{\rho}_{mn\bm{R}\bm{q}}(\bm{G})$ gives physically correct $R$ dependence of the Coulomb parameters as shown in Fig.~\ref{fig:V_Rdep}.


\add{In this study we do not study the effect of interlayer atomic diffusion, which makes the structure of the interface slightly obscured~\cite{Logvenov}.}

\subsection{Effect of structural optimization}

\begin{figure}
  \centering
  \includegraphics[width=8.0cm]{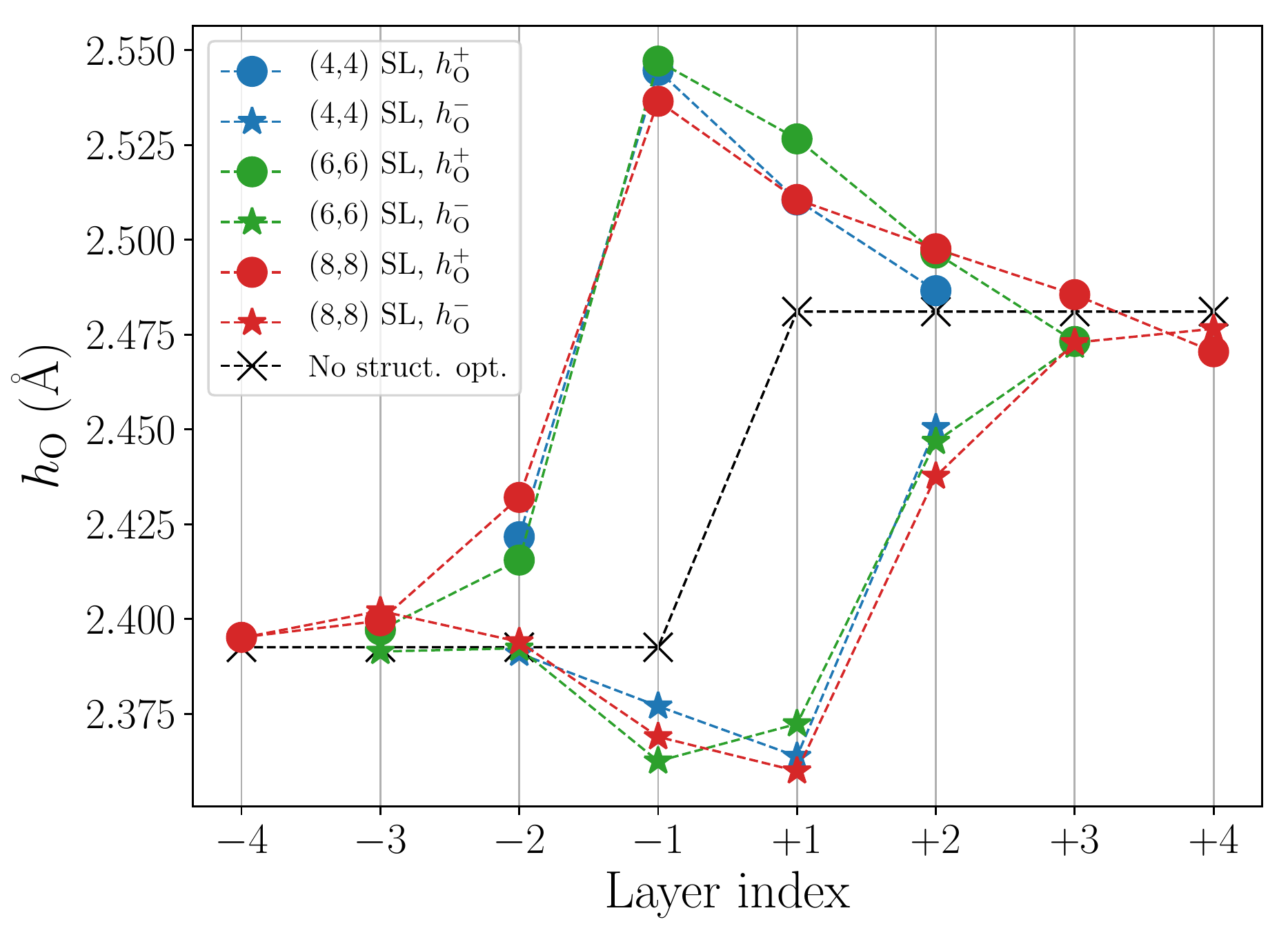}
  \caption{(color online) Layer dependence of the apical oxygen height $h_{\mathrm{O}}$ calculated for the three different sizes of the superlattice. The definitions of $h_{\mathrm{O}}^{+}$ and $h_{\mathrm{O}}^{-}$ are given in Fig.~\ref{fig:superlattice}. \add{The apical oxygen heights after the optimization are compared with those without the optimization (cross points).}}
  \label{fig:ho}
\end{figure}

The formation of an LCO/LSCO interface introduces an abrupt change of the electrostatic potential near the interface, which is energetically unfavorable. Therefore, the internal coordinates can deviate from the bulk values, which is likely to influence the DFT band structures and the associated $E_{g}$ Hamiltonian. 

Figure \ref{fig:ho} shows the layer dependence of the apical oxygen height obtained after performing the structural optimization. Since the mirror plane symmetry of the CuO$_{2}$ planes is lost due to the interface, the $h_{\mathrm{O}}$ values of the apical oxygen atoms above and below a CuO$_{2}$ plane are different from each other. To distinguish these two, we use  $h_{\mathrm{O}}^{\pm}$ as defined in Fig.~\ref{fig:superlattice}. After the relaxation, the $h_{\mathrm{O}}^{+}$ value, which is the distance to the apical oxygen on the LCO side, tends to increase from the bulk value, whereas $h_{\mathrm{O}}^{-}$ shows the opposite tendency. The difference $|h_{\mathrm{O}}^{+}-h_{\mathrm{O}}^{-}|$ becomes significant at the interface (layers $\pm 1$) and sharply decreases as going away from the interface. These structural changes were commonly observed in the studied SLs with different sizes. 

\begin{figure}
  \centering
  \includegraphics[width=7cm]{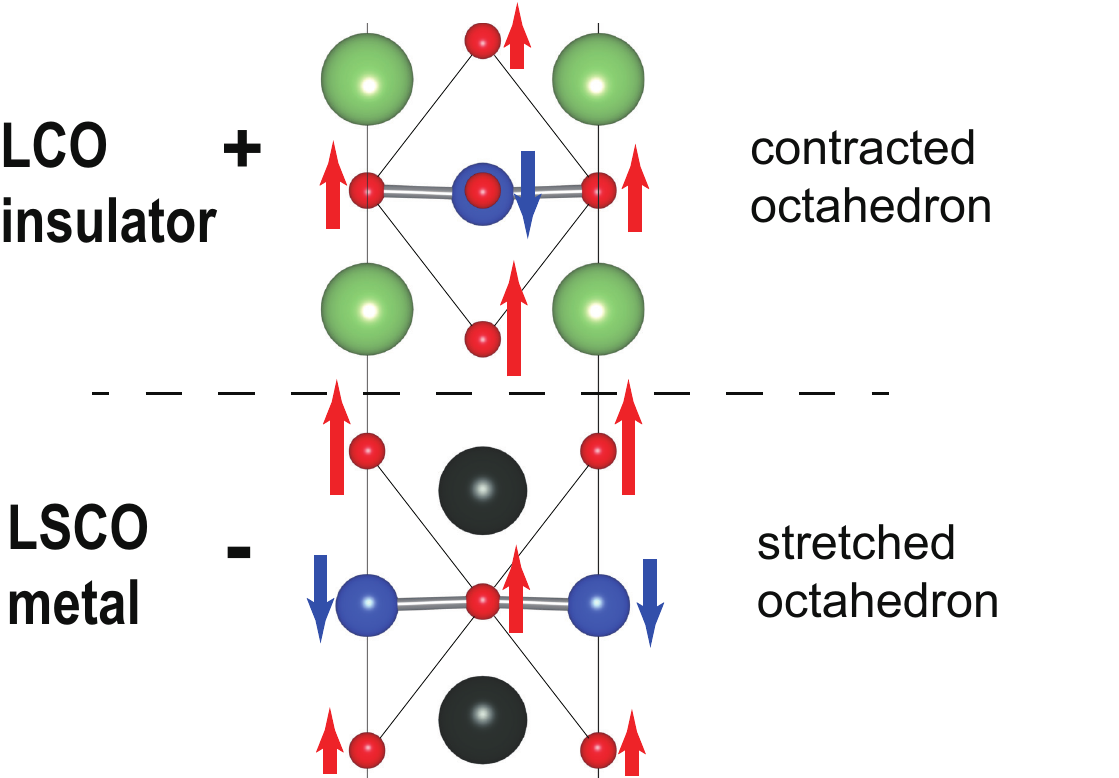}
  \caption{(color online) Distortion of octahedrons at the interface. The upward shift (directed to the insulator) of apical oxygens closest to the interface is the largest and the distortions become smaller when oxygen atom position becomes far from the interface. The copper atoms moves downward. As a result, the octahedrons in the insulator side but near the interface is contracted, while they are stretched in the LSCO side.}
  \label{fig:distortion}
\end{figure}

All of these behaviors can be understood qualitatively from the Mardelung potential. 
The ionic charge of La and Sr near the interface generates electric field from
 the LCO side to 
 the LSCO side \tb{at the interface, because the CuO$_2$ planes closest to the interface, one in the LCO side and the other in the LSCO side, are separated by a LaO plane in the LCO side and a (La,Sr)O plane in the LSCO side. Namely, because of the charge imbalance between the LaO and (La,Sr)O planes, it generates an electric field in the direction from the LSCO to the LCO sides, which causes the upward shift of the oxygen and downward shift of the copper atoms in the configuration illustrated in Fig.~\ref{fig:distortion}. This electric field also induces the electron transfer from the LCO side to the LSCO side and the electron itinerancy (interlayer electron hopping) makes its transfer range vertical to the interface wider. It also makes a wider range of atomic-position shift in a self-consistent fashion.
Since the electric field is strongest at the interface, the distortion is of course largest at the interface and becomes smaller at points far from the interface.}

The inversion symmetry breaking due to the structure relaxation induces a stretch (contraction) of the octahedron along $c$ axis in the LSCO (LCO) side  (Fig.~\ref{fig:distortion}), whose effect is discussed in detail in Sec. \ref{Sec:discussion}.
In addition, in plane Cu and O are not aligned in plane any more because of anti-phase distortion between Cu and O ions. 
The deviation of the O-Cu-O angle from 180$^\circ$ amounts to $\sim$4$^{\circ}$ at the interface. 

\begin{figure}[t]
  \centering
  \includegraphics[width=8.5cm]{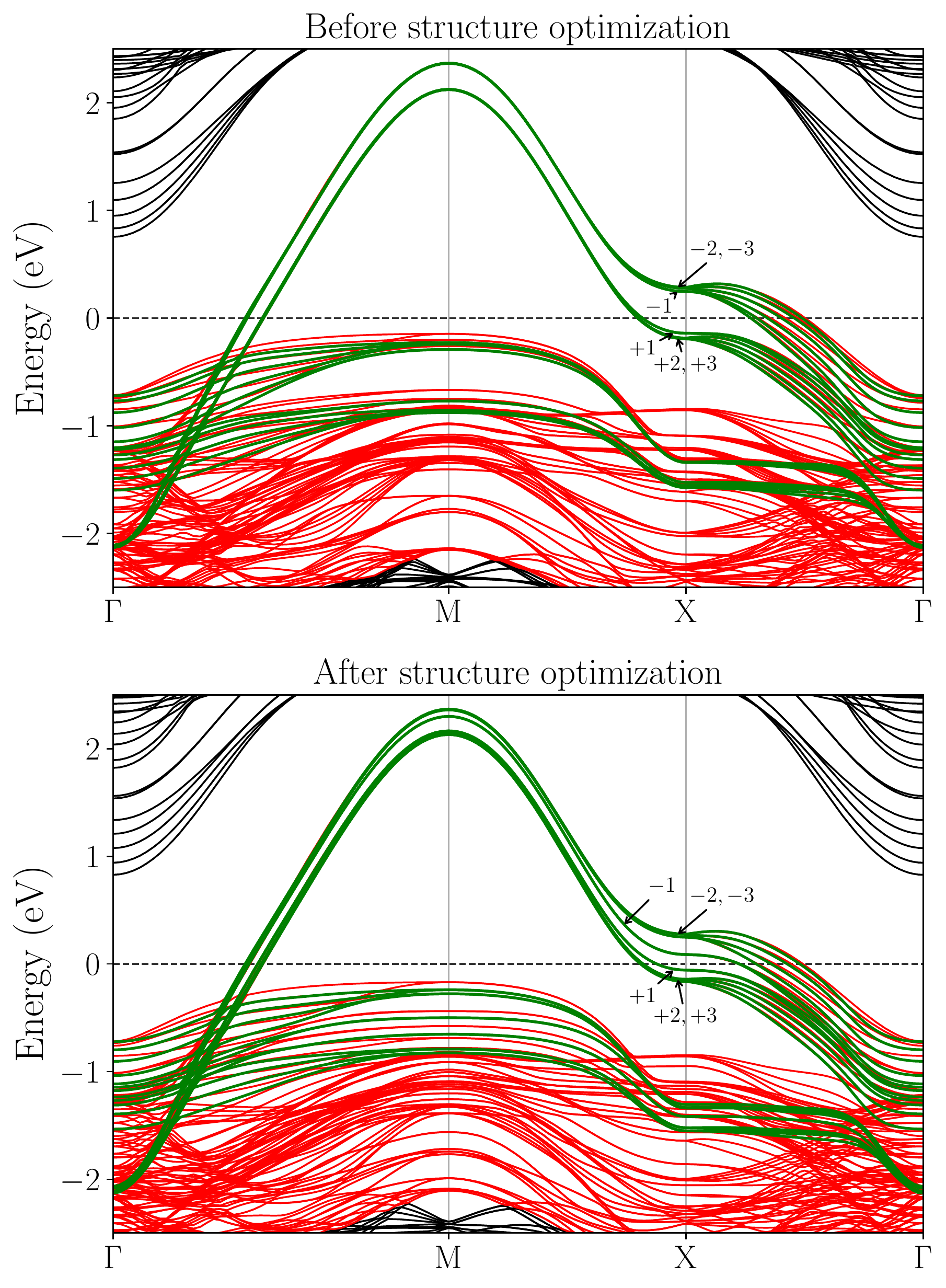}
  \caption{(color online) KS and Wannier band structures of the $(6,6)$ SL before (top panel) and after (bottom panel) structure optimization. For the notations and definitions of the lines, see Fig.~\ref{fig:bulk_band} caption.}
  \label{fig:DFT_band66}
\end{figure}

\begin{figure*}[t]
  \centering
  \includegraphics[width=\textwidth]{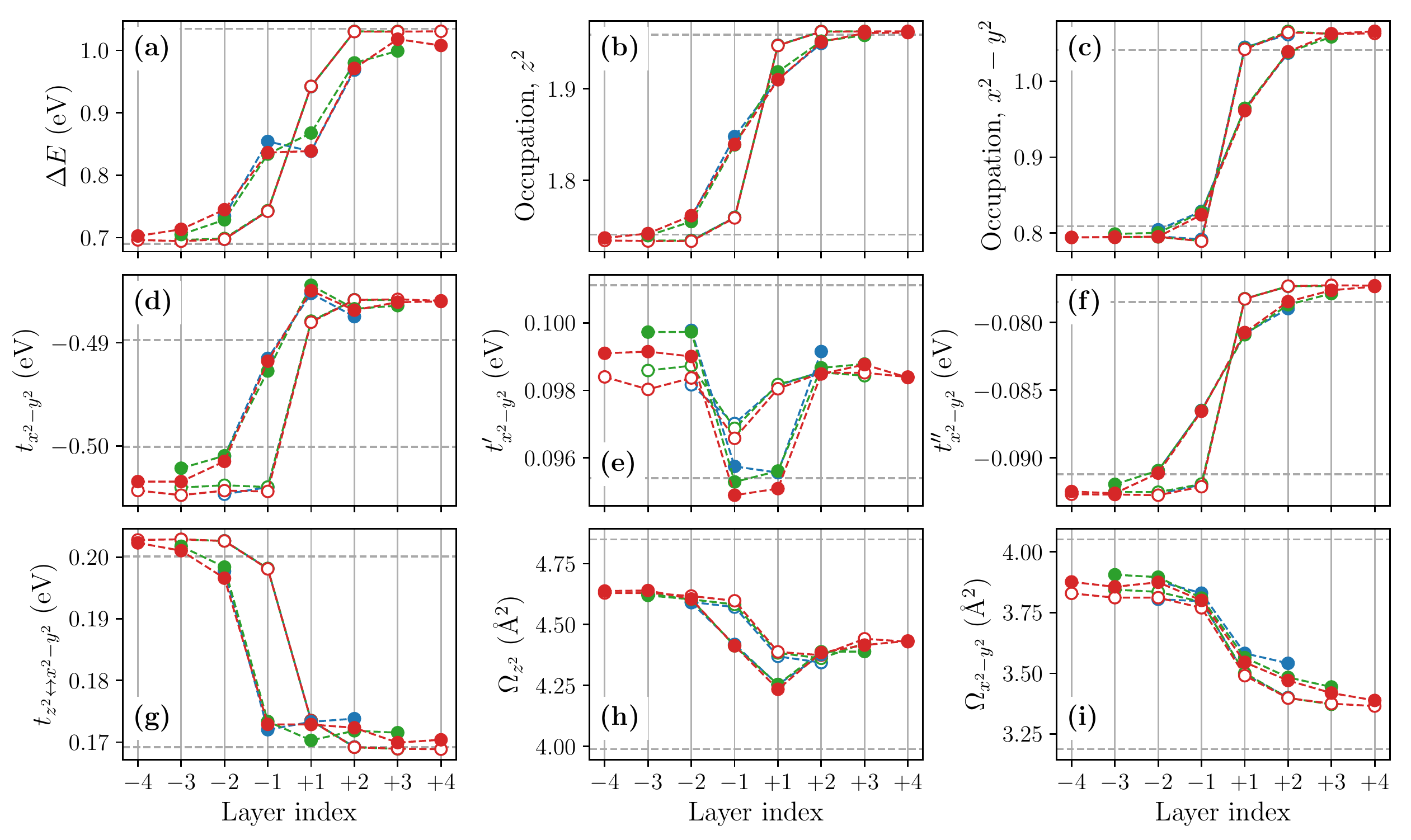}
  \caption{(color online) Layer dependence of (a) the level offset $\Delta E$, (b, c) the occupation number within PBE, (d, e, f, g) the hopping parameters related to the $d_{x^{2}-y^{2}}$ orbital, and (h, i) the spread of the MLWF calculated with the three different SL sizes. The blue, green and red symbols are results obtained for the $(4,4)$, $(6,6)$ and $(8,8)$ SLs, respectively. The open circles represent the data for the unrelaxed structures, while the filled circles are obtained after structure optimization. The horizontal dashed lines indicate the corresponding values of the bulk LCO and LSCO (N2 and D2 in Table~\ref{table:bulk_parameter}).}
  \label{fig:hopping_SL}
\end{figure*}

The structural relaxation considerably affects the electronic structure of the heterostructure as evidenced in Fig.~\ref{fig:DFT_band66}. For example, the orbital energies of the $d_{x^{2}-y^{2}}$-like orbital along the line M--X change very sharply at the interface before the structure optimization, in accord with the previous DFT calculation~\cite{Misawa2016SciAdv}. After the optimization, the orbital-energy shift occurs and the energy change becomes more gradual.
The shift of the energy-level is noticeable in the layers $\pm 1$, particularly around the point X, but it is far smaller in the other layers ($\pm 2$, $\pm 3$). This behavior is consistent with the rather strong deformation of the O-Cu-O angle in the layers $\pm 1$ and its rapid recovery in the layers $\pm 2$ observed in Fig.~\ref{fig:ho}. Therefore, these results indicate that the structure optimization influences the hopping parameters mainly at the layers $\pm 1$, which is investigated in the subsequent section.

\subsection{Layer-dependent hopping and Coulomb parameters}

Figure \ref{fig:hopping_SL} shows the layer dependence of the level offset $\Delta E$, the occupation number $n_{m} = \braket{a_{m\bm{0}}^{\dagger}a_{m\bm{0}}}$, the dominant part of the hopping parameters and the spread of the MLWFs calculated for the $(4,4)$, $(6,6)$ and $(8,8)$ SLs. To see the effect of the structure optimization, we compare the results before (open symbols) and after (filled symbols) the optimization in the figure. 

We observe that the structural optimization affects all of the parameters especially at the layers $\pm 1$, leading to a more gradual layer dependence. 
The change of the slope is particularly significant in the layer dependence of the level offset $\Delta E$ [Fig.~\ref{fig:hopping_SL}(a)], hole concentration given by $3-n_{z^2}-n_{x^2-y^2}$ [Figs.~\ref{fig:hopping_SL} (b) and (c)], and the interband hopping [Fig.~\ref{fig:hopping_SL} (g)].
For example, the hole concentration at the layer $+1$ changes from 0.01 to 0.13 by the structure optimization. The latter is close to an optimal doping level $x=0.15$ of the bulk \LSCOx at which a maximum  $T_{\mathrm{c}}$ has been observed. 


\begin{figure*}[t]
  \centering
  \includegraphics[width=\textwidth]{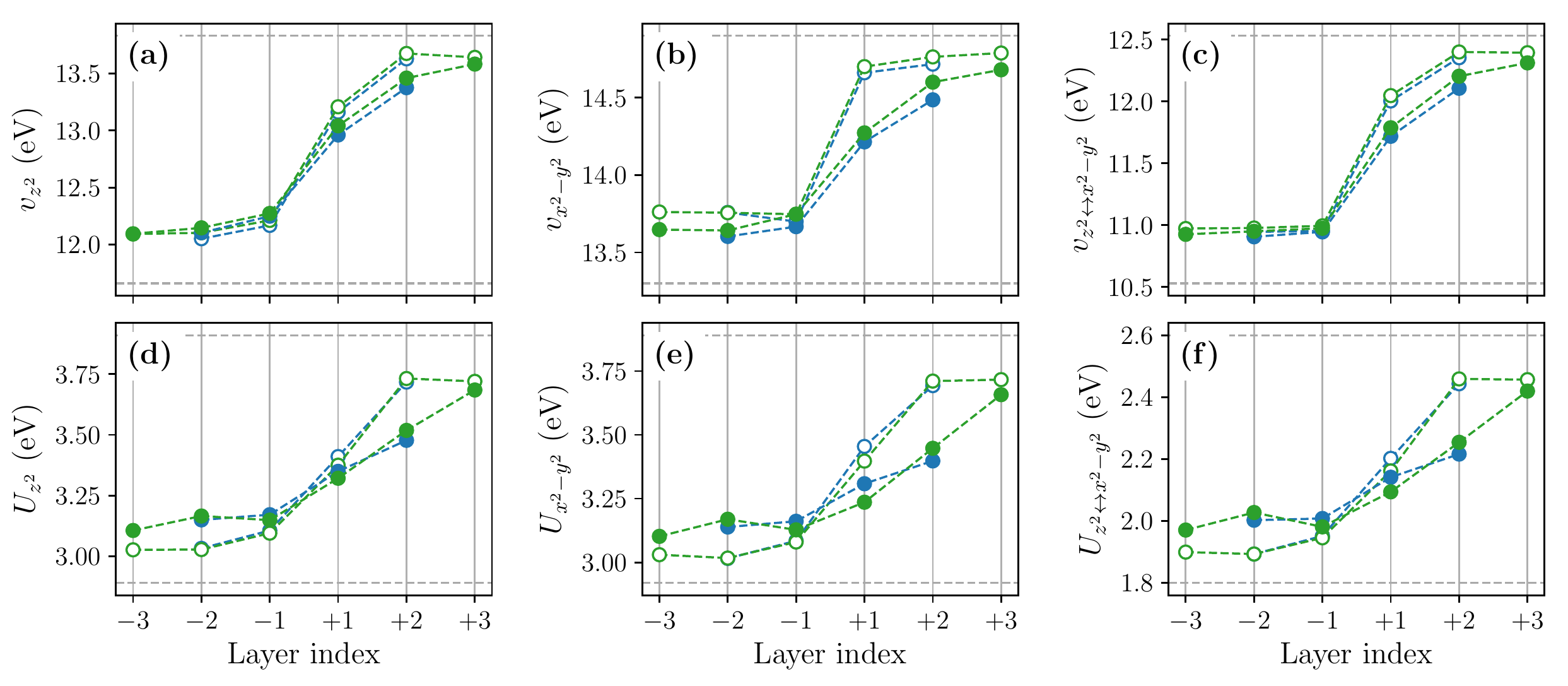}
  \caption{(color online) Layer dependence of  the bare (a, b, c) and partially screened (d, e, f) on-site Coulomb parameters calculated with the $(4,4)$ (blue symbols) and $(6,6)$ (green symbols) SLs. The open and filled symbols represent the result obtained for the unrelaxed and relaxed structures, respectively. The horizontal dashed lines indicate the corresponding values of the bulk LCO and LSCO (N2 and D2 in Table~\ref{table:bulk_parameter}).}
  \label{fig:Coulomb_SL}
\end{figure*}

The layer dependence of the onsite Coulomb parameters calculated for the $(4,4)$ and $(6,6)$ SLs are shown in Fig.~\ref{fig:Coulomb_SL}. 
As in the case of bulk systems discussed in Sec.~\ref{Sec:result_bulk}, the 
effective Coulomb interactions tend to be larger in the LCO side, which can be attributed to the smaller Wannier spread and the weaker screening.
The structure relaxation makes the layer dependence of interaction parameters gradual.  
As a result, at the layer $+1$, the $U$ value of the $d_{x^{2}-y^{2}}$ orbital is 3.24 eV, which is about 15\% smaller than the bulk LCO value.  

We also calculated the layer dependence of the off-site Coulomb $V_{n}$ and exchange $J_{x^{2}-y^{2}\bm{0};z^{2}\bm{0}}$ interactions (see Fig.~S1 of the SM). We observe that the layer dependence of the off-site Coulomb parameter $V_{n}$ is weak; $V_{x^{2}-y^{2}}$ and  $V_{z^{2}}$ values change in the range of 0.84--0.89 eV and 0.65--0.67 eV, respectively. By contrast, the layer dependence of the exchange parameter $J_{x^{2}-y^{2}\bm{0};z^{2}\bm{0}}$ is rather significant, which changes from 0.47 eV at the layer $-3$ to 0.39 eV at the layer $+3$. These tendencies are consistent with the doping-level dependence of $V_{n}$ and $J$ observed in the bulk systems (Table \ref{table:bulk_parameter}). 

\begin{figure}[t]
  \centering
  \includegraphics[width=6.5cm]{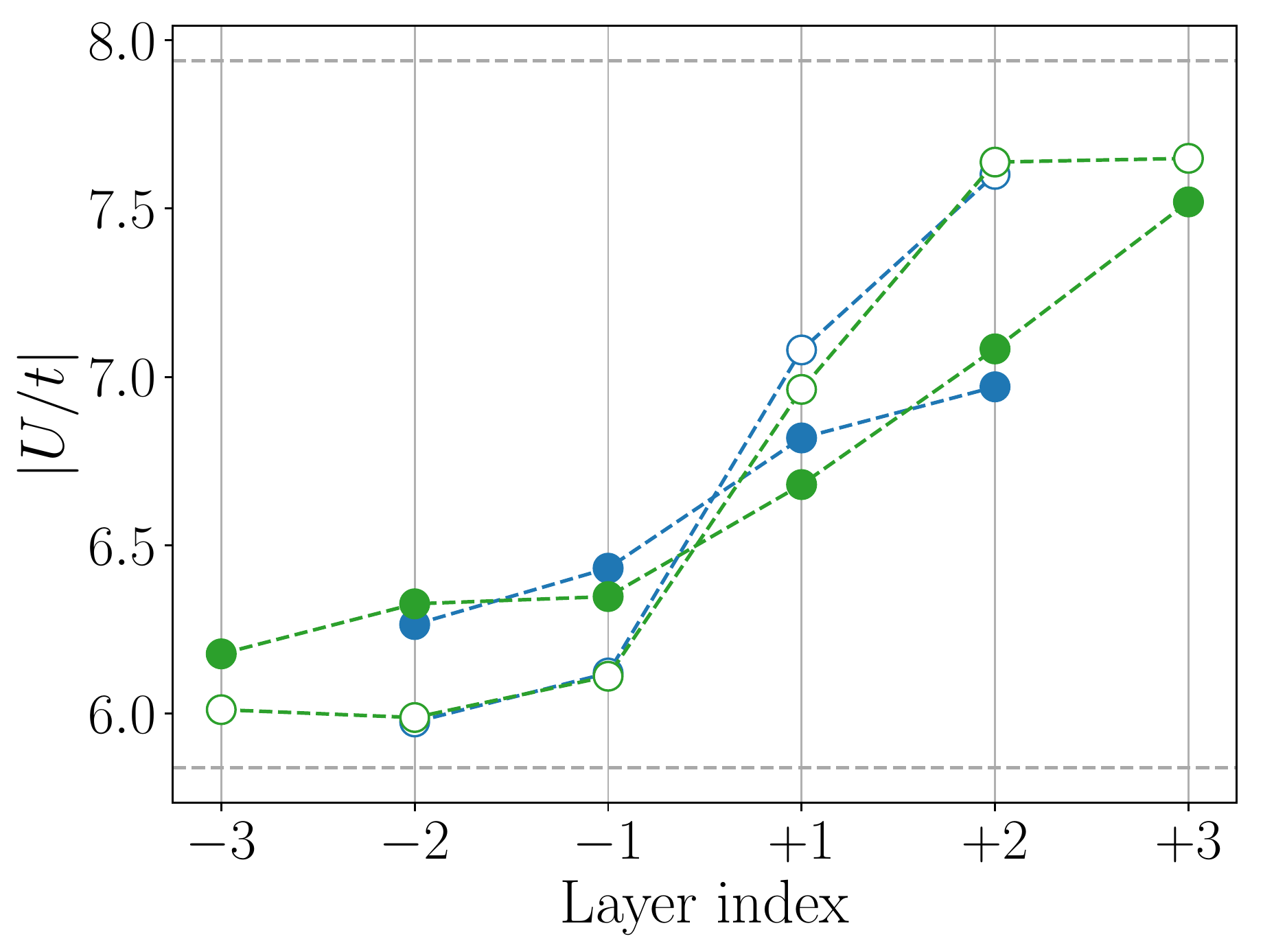}
  \caption{(color online) Layer dependence of scaled correlation strength $|U/t|$ calculated for the $(4,4)$ (blue symbols) and $(6,6)$ (green symbols) SLs before (open symbols) and after (filled symbols) structure optimization.}
  \label{fig:Ut_SL}
\end{figure}

Figure \ref{fig:Ut_SL} shows the layer dependence of the scaled correlation strength $|U/t|$ calculated for the $d_{x^{2}-y^{2}}$ orbital. Near the LCO/LSCO interface, the $|U/t|$ value decreases from the bulk LCO value of 7.94 to 6.68 at the layer $+1$, which amounts to a 15\% reduction. Since the relative stability of the superconducting phase over other competing phases changes rather sensitively with the $|U/t|$ value, considering the 15\% reduction of $|U/t|$ would be necessary to explain the unique properties of the superconductivity at interfaces quantitatively.

Finally, we discuss the convergence of parameters to bulk values.
We see that the calculated Coulomb parameters and the Wannier spread of the SLs did not reach the bulk LCO and LSCO values even at the layers farthest from the interface, indicating rather strong sensitivity of these parameters to the Madelung potential difference induced by the interface. This issue is expected to be resolved by using a much larger SL, which was not pursued in this study owing to the computational limitations.
Notwithstanding, since the two different sizes of the SL calculation shows more or less the same behaviors, the Coulomb parameters near the interface are likely to be already converged and therefore reliable enough for studying the superconductivity at the LCO/LSCO interface by low-energy solvers.

\section{Discussion}
\label{Sec:discussion}

Near the interface, $\Delta E$, transfers, hole concentrations and the interactions all show substantially more gradual change with moving from the LSCO side to the LCO side than the unrelaxed case. We here discuss that all the above characteristic features are 
explained by a basic principle ``the nature relaxes to avoid discontinuous changes". This principle is manifested concretely in the real material in the following way. 

Before the structural relaxation, the electrostatic potential of the LaO layer changes abruptly, and it generates rather strong electric field at the interface.
However, the shift of the atomic position and the electronic charge redistribution described in Sec. IVA leads to more gradual transition of the electron concentration between the LCO and LSCO sides. 
Since the negatively-charged apex oxygen approaches (moves away from) the CuO$_{2}$ layer on the LCO (LSCO) side on average as is illustrated in Fig.~\ref{fig:distortion}, 
the electronic level at the CuO$_{2}$ layer is raised (lowered) in the LCO (LSCO) side in comparison to the unrelaxed lattice.
It enhances  the electron transfer from the LCO side to the LSCO side originally caused by the electric field at the interfacial charge imbalance (see Sec. IVA) as is seen in Fig.~\ref{fig:hopping_SL}(b), (c).
Since the onsite level of $d_{x^{2}-y^{2}}$-like orbital is higher than that of $d_{z^2}$-like orbital,
the electron density decreases mainly from the $d_{x^{2}-y^{2}}$-like orbital in the LCO side and increases mainly in the the $d_{z^2}$-like orbital in the LSCO side as shown in Figs.~\ref{fig:hopping_SL}(b) and (c).

The level offset between the $d_{x^{2}-y^{2}}$-like and $d_{z^2}$-like orbitals, $\Delta E$, is determined mainly by the ligand field of the six oxygen ions surrounding Cu.
The contraction (stretch) of the octahedron for the LCO (LSCO) side leads to the decrease (increase) of the average distance to the apical O from Cu, $(h_{\mathrm{O}}^{+}+h_{\mathrm{O}}^{-})/2$, in the LCO (LSCO) side (see Fig.~\ref{fig:ho}). 
Because the onsite level of $d_{z^2}$ orbital is more sensitive to $h_{\mathrm{O}}$ than $d_{x^{2}-y^{2}}$ orbital, 
a larger $h_{\mathrm{O}}$ value leads to a larger $\Delta E$. 
Therefore, $\Delta E$ decreases (increases) on the LCO (LSCO) side after the structure relaxation as shown in Fig.~\ref{fig:hopping_SL}(a).

The nearest neighbor transfer $t_{x^2-y^2}$ relatively decreases from the unrelaxed lattice value  [Fig.~\ref{fig:hopping_SL}(d)] despite a slight increase in the Wannier spread [Fig.~\ref{fig:hopping_SL}(i)]. This is presumably because the inplane Cu and O do not align in the flat plane any more but form a zigzag alignment after the structural optimization.

The screened interaction of the antibonding $d_{x^{2}-y^{2}}$-like and $d_{z^{2}}$-like electrons decreases (increases) on the LCO (LSCO) side in comparison to the unrelaxed case. The structure relaxation increases (decreases) the hole concentration on the LCO (LSCO) side.
As we see in Sec.~\ref{Sec:bulk_doping_strain_dep}, larger hole concentration enhances the screening. 
Therefore increase (decrease) of the hole concentration may explain the reduction (enhancement) of $U$ values on the LCO (LSCO) side. 

In total, the octahedron distortion leads to more gradual layer dependence of all the quantities than the unrelaxed case following the above principle.Such weakened and more gradual layer dependence than that before lattice relaxation seems to follow a general principle: When the spatial gradient becomes strong, the system reacts to weaken it by screening such strong spatial dependence in analogy to Le Chatelier's priciple in the time dependence. Such screening effects may also work to weaken the effect of impurities, randomness and interface roughness in real interface.

More gradual layer dependence of the onsite level than that before the lattice relaxation may play a role to stabilize the superconductivity. 
In bulk compounds, there are strong tendency towards the in-plane charge inhomogeneity which suppresses superconductivity~\cite{zheng:2017,ido:2018,darmawan:2018}.  
However, the interface systems might be able to avoid the in-plane inhomogeneity by making out-of-plane inhomogeneity 
(interlayer phase separation).
In Ref.~\cite{Misawa2016SciAdv}, it has been argued that the gradual layer dependence of the onsite level is favorable to realize interlayer phase separation and stabilize superconductivity. 
It is indeed interesting to analyze the derived interface Hamiltonian 
to study the stability of the superconductivity. 

Finally, the sensitivity of onsite-level, hopping and Coulomb interaction parameters to the lattice distortion suggests nonnegligible coupling between electron and lattice degrees of freedom.
Therefore, it is also interesting to investigate the role of electron-phonon couplings in the interface, which might play a role to enhance superconductivity as suggested in other interface systems such as FeSe on SrTiO$_3$ substrate~\cite{doi:10.1146/annurev-conmatphys-031016-025242}.

\section{Conclusion}
\label{Sec:conclusion}

We have derived the \textit{ab initio} $E_{g}$ Hamiltonian of \LCO/\LSCO superlattices by performing large-scale cRPA calculations based on DFT. We have shown that the level offset between the $d_{x^{2}-y^{2}}$ and $d_{z^{2}}$ orbitals, $\Delta E$, and the Coulomb parameters near the interface become smaller than those of the LCO bulk values. This occurs even without performing structure relaxation and can be attributed to the change of the electrostatic potential induced by substituting La with Sr, which increase the energy level of the oxygen $2p$ orbitals more significantly than that of Cu $3d$ orbitals. After the structural relaxation, the layer dependence of the hopping and Coulomb parameters becomes more gradual than the unrelaxed case, which results from the contraction and stretch of CuO$_{6}$ octahedron in the LCO and LSCO side, respectively. The effect of the structural relaxation is particularly noticeable in $\Delta E$ and the occupation number. Since these parameters as well as the scaled correlation strength $|U/t|$  influence the stability of the superconducting state, the modulation of these parameters at the interface reported in this work should be considered for developing robust understandings of the unique superconducting properties observed in interfaces.

\acknowledgements

The authors thank Kazuma Nakamura, Motoaki Hirayama, Takahiro Ohgoe, Takahiro Misawa, Youhei Yamaji, Kota Ido, and Kosuke Miyatani for fruitful discussion.
This work was partially supported by the MEXT HPCI Strategic Programs, and the Creation of New Functional Devices and High-Performance Materials to Support Next Generation Industries (CDMSI) and a Grant-in-Aid for Scientific Research (No. 16H06345) from Ministry of Education, Culture, Sports, Science and Technology, Japan. The computation in this work has been done using the facilities of the Supercomputer Center, Institute for Solid State Physics, The University of Tokyo.
Y. N. was supported by Grant-in-Aids for Scientific Research (JSPS KAKENHI) (No. 17K14336 and 18H01158).

\appendix

\section{Comparison of the present cRPA calculation for bulk \LCO with previous works}
\label{Sec:cRPA_comparison}

In the original disentanglement method of Şaşıoğlu \textit{et al.}~\cite{Sasoglu:2011ch}, the $d$-$d$ polarization process is excluded via the weight of the $E_g$ subspace $w_{\alpha\bm{k}}$. Therefore, the screening process involving the $d_{x^{2}-y^{2}}$-like KS orbitals crossing the Fermi energy can be avoided perfectly if the $w_{\alpha\bm{k}}$ values become exactly one for these KS orbitals. For \LSCOx, however, we observed that the weight slightly deviates from one even near the Fermi energy when the KS energy $\epsilon_{\alpha\bm{k}}$ is outside the frozen window. For example, the weight of the $d_{x^{2}-y^{2}}$-like KS orbitals at $\bm{k} = (\frac{\pi}{4},0,0)$ was around 0.97, and the remaining weight of 0.03 origites from the orbitals outside the $E_{g}$ subspace, which contribute to $P_{r}$. Since we want to avoid such a small screening process from the $d_{x^{2}-y^{2}}$-like KS orbitals completely as the disentanglement method of Miyake \textit{et al.}~\cite{Miyake:2009kx} does, we updated the $w_{\alpha\bm{k}}$ values in such a way that those close to one (zero) becomes exactly one (zero) while keeping the total weight $\sum_{\alpha}w_{\alpha\bm{k}}$ unaltered. Such a treatment did not change the results significantly but increased the partially-screened Coulomb parameters by $\sim$5\% compared with the original treatment, leading to better agreement with the previous result of $U_{x^{2}-y^{2}} = 4.2$ eV~\cite{miyatani} which was obtained based on the full potential linearized muffin tin orbital (FP-LMTO) method and the disentanglement method of Miyake and coauthors. The present result of $U_{x^{2}-y^{2}} \sim 3.9$ eV (3.67 eV with 500 bands) is still smaller than 4.2 eV, which can most likely be attributed to the difference in the Wannierization parameters and/or the pseudopotentail adopted in this study. Indeed, the bare Coulomb parameters in this study is also smaller than the FP-LMTO based study by $\sim$15\%.

In the previous cRPA studies of LCO, the $U_{x^{2}-y^{2}}$ values of 3.65 eV~\cite{PhysRevB.91.125142} and 3.15 eV~\cite{Jang:2016aa} have been reported, which agree reasonably well with our result especially given that the $U$ value is rather sensitive to the detail of the Wannierization procedure and the resulting spread of the Wannier orbital. 
The result of Ref.~\cite{PhysRevB.91.125142} was obtained for the one-band Hamiltonian, where the Wannier function was constructed without the frozen window near the fermi energy. Therefore, the shape (spread) of their $d_{x^{2}-y^{2}}$-like Wannier orbital is similar to ours (Fig.~\ref{fig:bulk_isosurface}(b)), resulting in the similar $U_{x^{2}-y^{2}}$ values. 
If the frozen window is used when constructing the one-band Hamiltonian, the resulting Wannier orbital should be more extended and the $U_{x^{2}-y^{2}}$ value should become smaller because of the strong hybridization between the $E_g$ orbitals in LCO. 
Compared to the results of Ref.~\cite{PhysRevB.91.125142} and ours, the $U_{x^{2}-y^{2}}$ value of Ref.~\cite{Jang:2016aa} seems somewhat smaller, whose origin is unclear due to the missing details of the Wannierization parameters in Ref.~\cite{Jang:2016aa}.

Recently, Hirayama \textit{et al.}~\cite{Hirayama:2018} has reported \textit{ab initio} effective Hamiltonians for bulk cuprates, including LCO, obtained within the FP-LMTO and the cGW method supplemented by the self-interaction correction (SIC) of the Hartree term~\cite{Hirayama:2013,Hirayama:2017}. The onsite Coulomb parameters of the $E_{g}$ Hamiltonian is reported to be $U_{x^{2}-y^{2}} = $ 5.3--5.5 eV, which is $\sim$25--30\% larger than the previous cRPA result of 4.2 eV~\cite{miyatani}. This enhancement in $U$ can be attributed to the refined treatment of the Coulomb interaction by the GW approximation, which makes the band width of the $d_{x^{2}-y^{2}}$-like orbital smaller and thereby increases the bare $V$ value by $\sim$5\%. More importantly, the GW calculation increases the level offset between the $d_{x^{2}-y^{2}}$-like orbital and other orbitals in the $r$ subspace, such as the bonding state formed by the Cu $d_{x^{2}-y^2}$ and in-plane O $p_{\sigma}$ orbitals, leading to the considerably weaker screening compared with cRPA.
While the cGW-SIC scheme is theoretically more refined than cRPA, it is computationally more demanding than cRPA and its application to the LCO/LSCO interface was impractical. Therefore, cRPA is used both for the bulk and heterostructure in this work, which still gives reasonably accurate results and does not change the conclusions of this paper which mainly focuses on the effects of the structural optimization on the $E_g$ Hamiltonians of the LCO/LSCO interface.

\section{Averaging method for Coulomb parameters}

\label{Sec:Averaging}

\begin{figure}[t]
  \centering
  \includegraphics[width=8.0cm, clip]{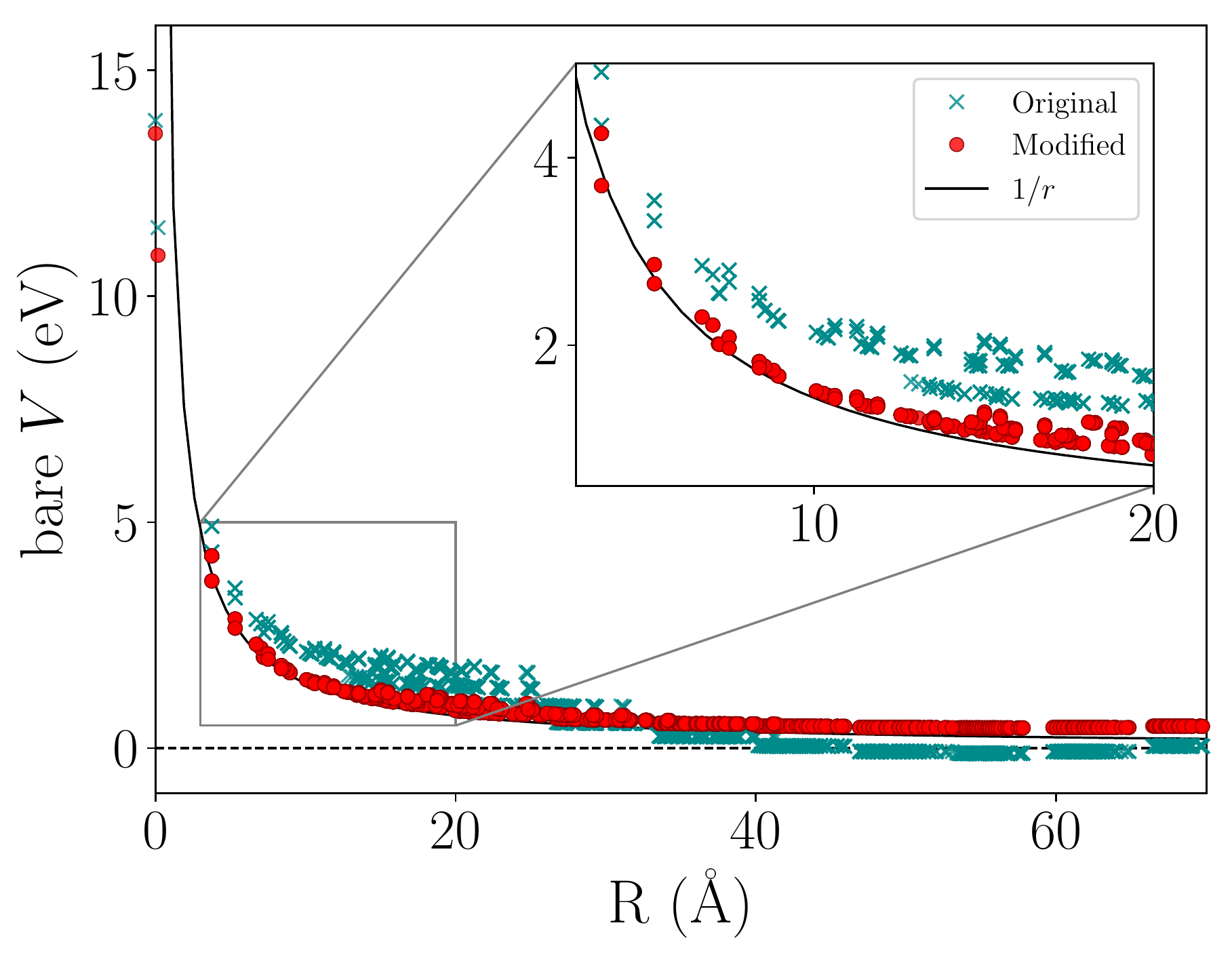}
  \caption{(color online) Bare Coulomb interaction $V_{m\bm{0}n\bm{R}}$ as a function of distance $|\bm{R}|$ calculated for the $(4,4)$ SL. The center of the $d_{x^{2}-y^{2}}$ Wannier orbital at the layer $-2$ is selected as the reference point $m\bm{0}$. The length of the $c$ axis of the the $(4,4)$ SL is 53.6~\AA{}, around which $V$ becomes negative when Eq.~(\ref{eq:U_orig}) is used. This issue is cured by the modified version Eq.~(\ref{eq:rho_avg}).}
  \label{fig:V_Rdep}
\end{figure}

In the present cRPA calculation, we employ the $8\times8\times2$ $\bm{k}$ points to use the tetrahedron method for an accurate numerical integration of Eq.~(\ref{eq:Pd}). However, we found that the original Eq.~(\ref{eq:U_orig}) gave unphysical negative $V$ values around $\bm{R} = (0,0,c)$ as shown in Fig.~\ref{fig:V_Rdep}. The negative contribution around  $\bm{R} = (0,0,c)$ comes mostly from the component of $\rho_{mn\bm{R}\bm{q}}(\bm{G})$ around $\bm{q} = (0,0,\frac{\pi}{c})$, which is far larger than the largest positive contributions from $\bm{q} = (\frac{\pi}{4a}, 0, 0)$ due to the prefactor $1/|\bm{q}|$ as well as the large anisotropy of the lattice shape, i.e., $a\ll c$. If one can increase the $\bm{k}$-mesh density up to $N\times N\times 2$ so that $\frac{\pi}{c} \simeq \frac{2\pi}{N a}$ is satisfied, the negative $v$ problem could be solved. However, such a calculation is almost infeasible because $N$ must be as large as 28 even for the smallest $(4,4)$ SL to satisfy the condition. 

To mitigate this issue, we simply modify the original $\rho_{mn\bm{R}\bm{q}}(\bm{G})$ given in Eq.~(\ref{eq:rho}) as 
\begin{multline}
  \tilde{\rho}_{mn\bm{R}\bm{q}}(\bm{G}) \approx \frac{1}{N|\bm{q}+\bm{G}|_{\mathrm{rep.}}} \\
  \times \sum_{\bm{k}}e^{-i\bm{k}\cdot\bm{R}} \braket{\psi_{m\bm{k}+\bm{q}}^{(w)}|e^{\mathrm{i}(\bm{q}+\bm{G})\cdot\bm{r}}|\psi_{n\bm{k}}^{(w)}}, \label{eq:rho_avg}
\end{multline}
where $|\bm{q}|_{\mathrm{rep.}}$ is a representative value of the norm around $\bm{q}$
defined as
\begin{equation}
  \frac{1}{|\bm{q}|_{\mathrm{rep.}}} = \left[\frac{1}{\Omega_{S_{\bm{q}}}}\int_{S_{\bm{q}}} \frac{1}{|\bm{k}|^{2}} d\bm{k}\right]^{\frac{1}{2}}.
\end{equation}
Here, $S_{\bm{q}}$ is the Wigner--Seitz cell of the lattice point $\bm{q}$, and $\Omega_{S_{\bm{q}}}$ is its volume.
This treatment assumes that a variation of $\braket{\psi_{m\bm{k}+\bm{q}}^{(w)}|e^{\mathrm{i}(\bm{q}+\bm{G})\cdot\bm{r}}|\psi_{n\bm{k}}^{(w)}}$ inside $S_{\bm{q}+\bm{G}}$ is small, which is satisfied when the size of $S_{\bm{q}}$ is reasonably small. If we use $\tilde{\rho}_{mn\bm{R}\bm{q}}(\bm{G})$, we can obtain a correct $1/r$ dependence without negative values as shown in Fig.~\ref{fig:V_Rdep}.

%
  
\widetext
\clearpage

\setcounter{equation}{0}
\setcounter{figure}{0}
\setcounter{table}{0}
\setcounter{section}{0}
\makeatletter
\renewcommand{\theequation}{S\arabic{equation}}
\renewcommand{\thefigure}{S\arabic{figure}}
\renewcommand{\thetable}{S\arabic{table}}
\renewcommand{\thesection}{S\arabic{section}}
\renewcommand{\bibnumfmt}[1]{[S#1]}
\renewcommand{\citenumfont}[1]{S#1}

\begin{center}
  \textbf{\large Supplemental Material}
\end{center}

\section*{S1. Layer dependence of off-site Coulomb and exchange parameters}

\begin{figure}[hb]
  \centering
  \includegraphics[width=0.8\textwidth]{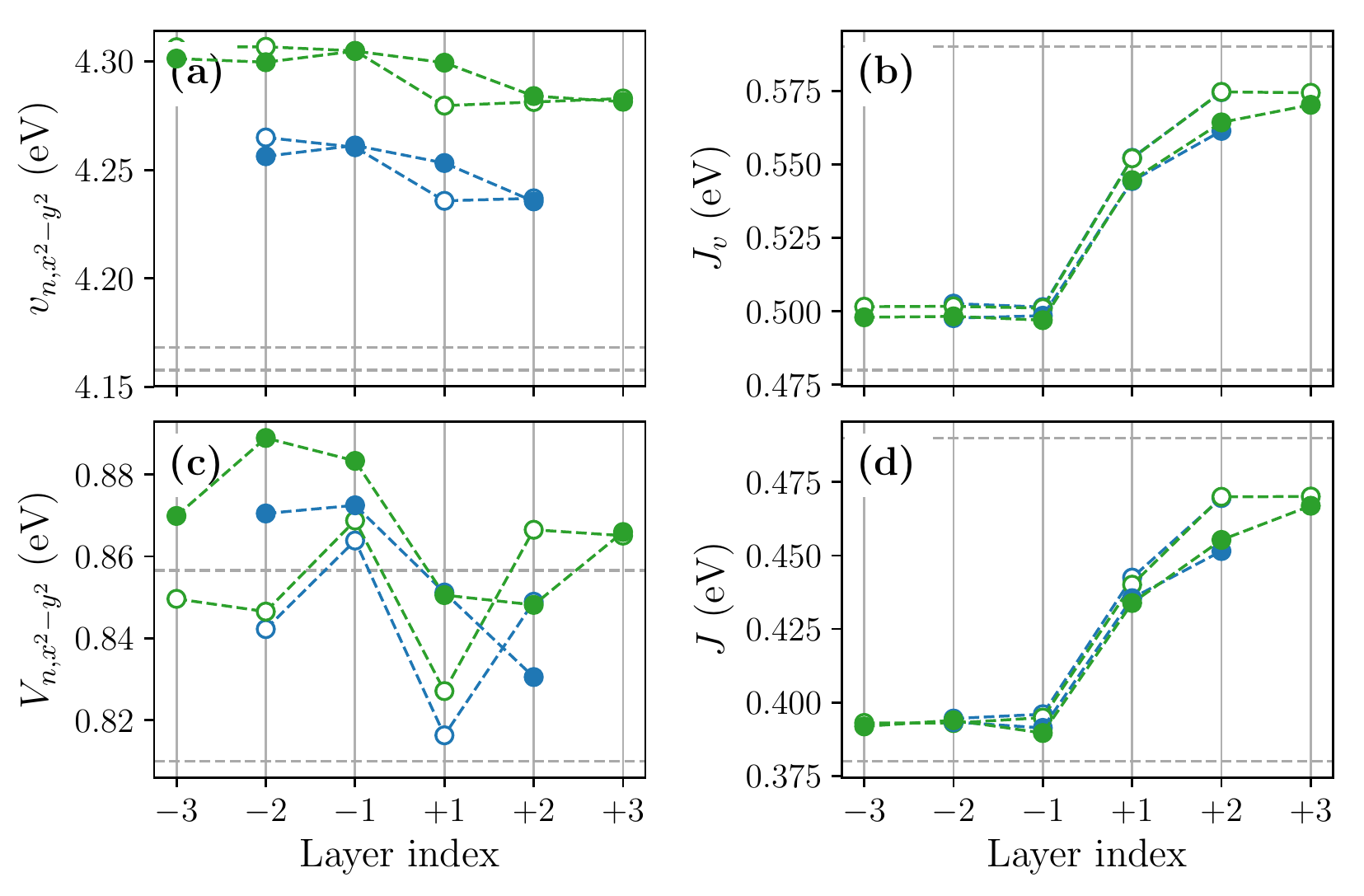}
  \caption{Layer dependence of the bare (a, b) and partially screened (d, e) off-site Coulomb and exchange parameters calculated with the $(4,4)$ and $(6,6)$ SLs after structure optimization. Here, the MLWFs are constructed with including 8 valence bands per CuO$_{2}$ layer in the outer window. The horizontal dashed lines indicate the corresponding values of the bulk LCO and LSCO.}
  \label{fig:Vn_SL_outer8}
\end{figure}

Figure \ref{fig:Vn_SL_outer8} shows the layer dependence of the off-site Coulomb and exchange parameters obtained with the same outer window setting as the main text, i.e., 8 valence bands per CuO$_{2}$ layer. In comparison with the on-site Coulomb parameters (Fig.~9 of the main text), the layer dependence of the off-site Coulomb parameters is weak as shown in Fig.~\ref{fig:Vn_SL_outer8} (a) and (c). The maximum difference of the $V_n$ values between the layers is $\sim$ 0.05 eV, which is smaller than 6\% of the average value. By contrast, the layer dependence of the exchange parameters is rather significant, and the maximum difference reaches $\sim$ 0.075 eV, which amounts to 15--19\% of the absolute $J$ values. The layer dependence of the exchange parameters are similar to that of the on-site Coulomb parameters.

\section*{S2. Parameters with wider outer window}

In the main text of the paper, we show the hopping and Coulomb parameters based on the rather extended $d_{z^{2}}$ orbital constructed from a narrower outer window that includes 8 valence bands per CuO$_{2}$ layer. Since the parameters of the low-energy effective Hamiltonian depend on the shape of the Wannier orbital, whose spread is sensitive to the outer window in \LSCOx, we also show the computational results obtained when we included 14 valence bands per CuO$_{2}$ layer in the outer window.

\begin{figure*}[t]
  \centering
  \includegraphics[width=\textwidth]{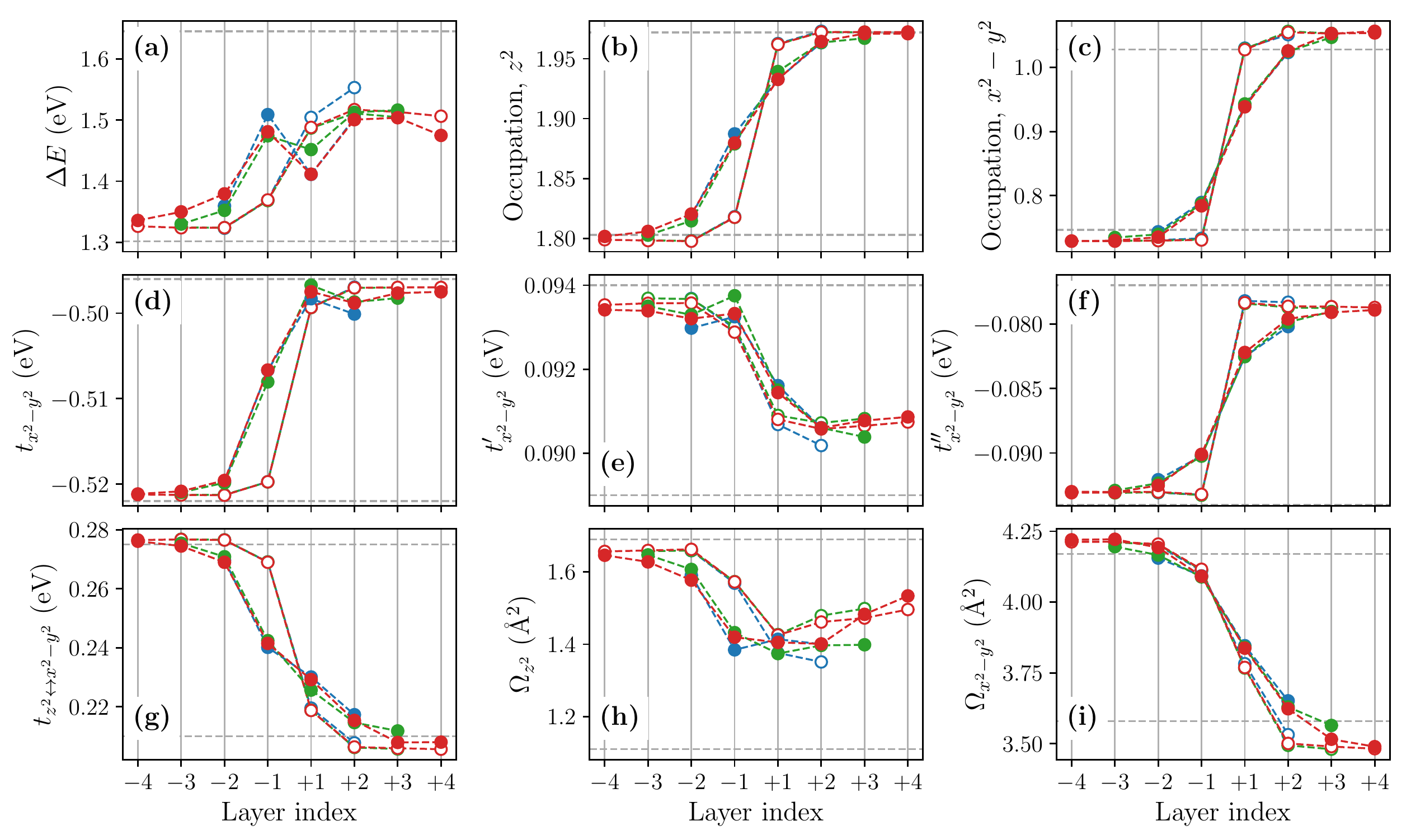}
  \caption{(color online) Layer dependence of (a) the level offset $\Delta E$, (b, c) the occupation number within PBE, (d, e, f, g) the hopping parameters related to the $d_{x^{2}-y^{2}}$ orbital, and (h, i) the spread of the MLWF calculated with the three different SL sizes. Here, the MLWFs are constructed with the wider outer window including 14 valence bands per CuO$_{2}$ layer. The blue, green and red symbols are results obtained for the $(4,4)$, $(6,6)$ and $(8,8)$ SLs, respectively. The open circles represent the data for the unrelaxed structures, while the filled circles are obtained after structure optimization. The horizontal dashed lines indicate the corresponding values of the bulk LCO and LSCO (N2 and D2 in Table 1 of the main text).}
  \label{fig:hopping_SL_outerwindow14}
\end{figure*}

\begin{figure*}[t]
  \centering
  \includegraphics[width=\textwidth]{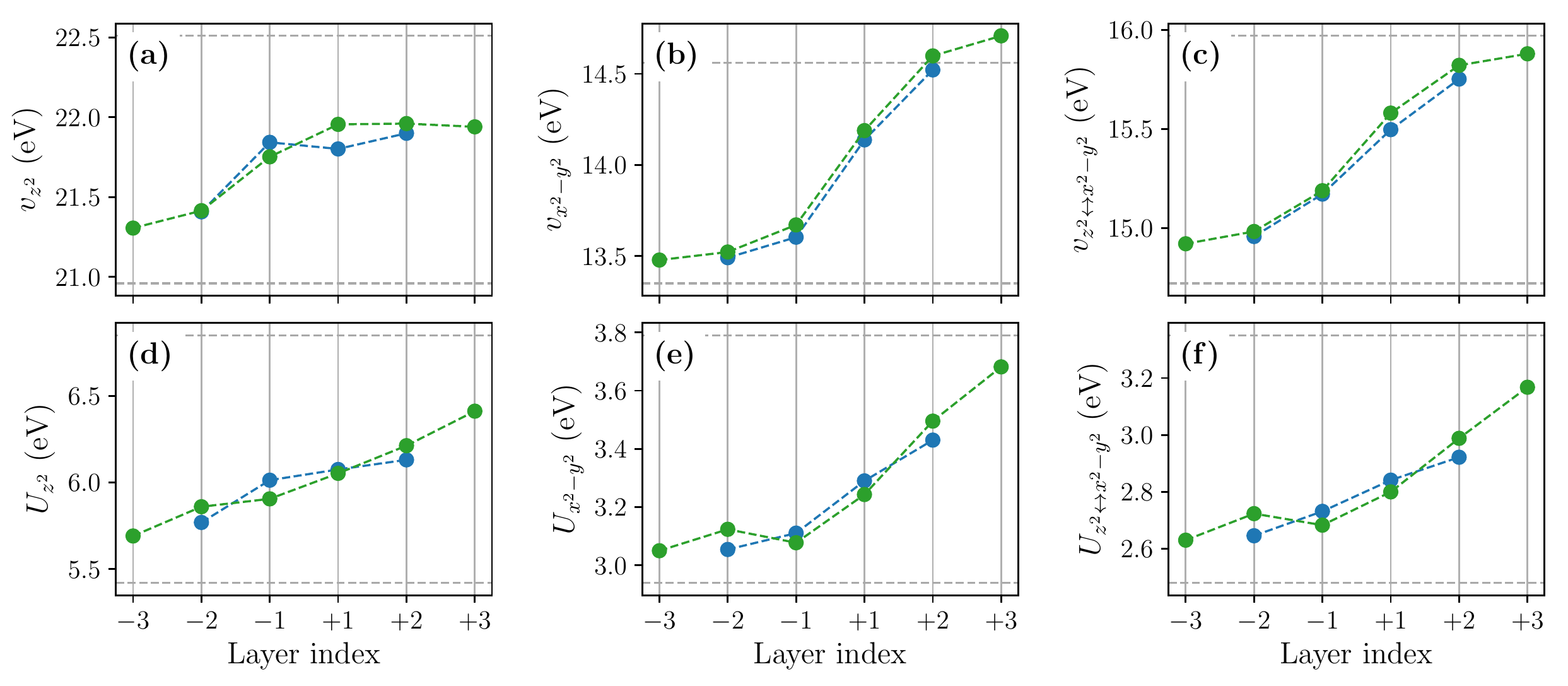}
  \caption{(color online) Layer dependence of  the bare (a, b, c) and partially screened (d, e, f) on-site Coulomb parameters calculated with the $(4,4)$ and $(6,6)$ SLs after structure optimization. Here, the MLWFs are constructed with the wider outer window including 14 valence bands per CuO$_{2}$ layer. The horizontal dashed lines indicate the corresponding values of the bulk LCO and LSCO (N2 and D2 in Table 1 of the main text).}
  \label{fig:Coulomb_SL_outer14}
\end{figure*}

Table \ref{table:bulk_parameter} shows the calculated hopping and Coulomb parameters of bulk \LSCOx, and Figs.~\ref{fig:hopping_SL_outerwindow14} and \ref{fig:Coulomb_SL_outer14} show the layer-dependent hopping and Coulomb parameters of the LCO/LSCO heterostructures, respectively. It can be inferred by comparing these data with Table 2, Figs. 8 and 9 of the main text that parameters involving the $d_{z^{2}}$ orbital changes dramatically with the change of the outer window. For example, the on-site Coulomb parameter of the bulk N2 system increases from 13.83 eV to 22.51 eV for the bare interaction $v$ and from 3.91 eV to 6.85 eV for $U(0)$. These enhancement can be attributed to the smaller Wannier spread of the $d_{z^{2}}$ orbital constructed from the 14 valence bands than that constructed from the 8 valence bands. The shrinkage of the Wannier orbital occurs due to the inclusion of the bonding state formed by the copper $d_{z^2}$ and the apical oxygen $p_z$ orbitals in the outer window. By contrast, the parameters that only involve the $d_{x^{2}-y^{2}}$ orbital such as $v_{x^{2}-y^{2}}$ and $U_{x^{2}-y^{2}}(0)$ do not change quantitatively because we employed the frozen window in the both outer-window settings.

\begin{table}

  \footnotesize
  \renewcommand{\baselinestretch}{1.0} 
  \caption{Hopping and Coulomb parameters of the bulk $E_{g}$ Hamiltonians calculated with different doping level $x$ and structural parameters.  Here, the MLWFs are constructed with the wider outer window including 14 valence bands per CuO$_{2}$ layer. The indices $m, n$ label the MLWFs; 0 and 1 correspond to  $d_{z^{2}}$ and $d_{x^{2}-y^{2}}$, respectively. The meanings of other parameters are the followings. $\Delta E$ : The on-site potential difference between the $d_{x^{2}-y^{2}}$ and $d_{z^{2}}$ orbitals; $t, t', t''$ : nearest, next-nearest, and second next-nearest hopping parameters, respectively; $v, U$: on-site Coulomb interactions; $v_{n}, V_{n}$: nearest-neighbor Coulomb interactions; $J_{v}, J$ : exchange interactions; $|U/t|$ : scaled correlation strength for the $d_{x^{2}-y^{2}}$ orbital.}
  \label{table:bulk_parameter}
  \begin{ruledtabular}
    \begin{tabular}{cccccc}
      \renewcommand{\arraystretch}{0.1}
      \setlength{\tabcolsep}{1pt}
                   &         & N1 ($x=0$) & N2 ($x=0$) & D1 ($x=0.45$) & D2 ($x=0.45$) \\ 
                   & $(m,n)$ &                   &                   &                      &                      \\ \hline
       $\Delta E$ 
                   &         &  $1.257$          & $ 1.645$          & $ 0.950$             & $ 1.302$  \\
               $t$ & $(0,0)$ & $-0.008$          & $-0.010$          & $-0.010$             & $-0.010$ \\
                   & $(0,1)$ & $ 0.214$          & $ 0.210$          & $ 0.271$             & $ 0.275$ \\
                   & $(1,1)$ & $-0.471$          & $-0.496$          & $-0.500$             & $-0.522$ \\
              $t'$ & $(0,0)$ & $-0.039$          & $-0.034$          & $-0.053$             & $-0.051$ \\
                   & $(0,1)$ & $ 0.000$          & $ 0.000$          & $ 0.000$             & $ 0.000$ \\
                   & $(1,1)$ & $ 0.088$          & $ 0.089$          & $ 0.093$             & $ 0.094$ \\
             $t''$ & $(0,0)$ & $-0.009$          & $-0.004$          & $-0.020$             & $-0.016$ \\
                   & $(0,1)$ & $ 0.034$          & $ 0.036$          & $ 0.041$             & $ 0.045$ \\
                   & $(1,1)$ & $-0.072$          & $-0.077$          & $-0.089$             & $-0.094$ \\
      Spread of the MLWF (\AA$^{2}$)
                   & $(0,0)$ & 1.37  &  1.11 &  2.09 &  1.69\\
                   & $(1,1)$ & 3.69  &  3.58 &  4.04 &  4.17 \\ \hline
      Bare Coulomb parameters \\
             $v$   & $(0,0)$ & 21.94 & 22.51 & 19.78 & 20.96\\ 
                   & $(0,1)$ & 15.77 & 15.97 & 14.46 & 14.72\\
                   & $(1,1)$ & 14.53 & 14.56 & 13.56 & 13.35\\
       $v_{\mathrm{n}}$
                   & $(0,0)$ & 3.65 & 3.73 & 3.60 & 3.69 \\
                   & $(0,1)$ & 3.86 & 3.93 & 3.85 & 3.92 \\
                   & $(1,1)$ & 4.04 & 4.10 & 4.08 & 4.12 \\
          $J_{v}$  & $(0,1)$ & 0.79 & 0.79 & 0.73 & 0.73 \\ \hline
      Partially screened Coulomb parameters \\    
           $U$  & $(0,0)$ & 6.67 & 6.85 & 5.21 & 5.42 \\
                   & $(0,1)$ & 3.34 & 3.35 & 2.56 & 2.48 \\
                   & $(1,1)$ & 3.82 & 3.79 & 3.11 & 2.94 \\
       $V_{\mathrm{n}}$ 
                   & $(0,0)$ & 0.73 & 0.75 & 0.69 & 0.69 \\
                   & $(0,1)$ & 0.78 & 0.79 & 0.76 & 0.75 \\
                   & $(1,1)$ & 0.84 & 0.84 & 0.83 & 0.80 \\
           $J$  & $(0,1)$ & 0.69 & 0.68 & 0.61 & 0.61 \\
           $|U/t|$ & $(1,1)$ & 8.11 & 7.64 & 6.22 & 5.63
    \end{tabular}    
  \end{ruledtabular}  
\end{table}

\end{document}